\documentclass[preprint,11pt]{elsarticle}

\usepackage{url}
\usepackage[hidelinks]{hyperref}
\hypersetup{pdfauthor=author}
\usepackage[margin=1in]{geometry}
\usepackage{bm}

\usepackage{amssymb}
\usepackage{amsmath}
\usepackage{subcaption}
\usepackage{tikz}
\usepackage{booktabs}
\usepackage{orcidlink}

\usepackage{lineno}

\begin{document}
\begin{frontmatter}

\title{Investigating Forecast Proficiency of Hurricane-Induced Compound Flooding With a Discontinuous Galerkin Shallow Water Equation Solver} 

\author[inst1]{Matthew Scarborough\corref{cor1}}
\ead{matthew.scarborough@nmbu.no}
\cortext[cor1]{Corresponding author \orcidlink{0009-0008-4667-9588}}
            
\author[inst2]{Chayanon Wichitrnithed}
\author[inst3,inst4]{Shintaro Bunya}
\author[inst5]{Ethan J. Kubatko}
\author[inst5]{Suranjan Nepal}
\author[inst2]{Clint Dawson}
\author[inst6,inst7]{Eirik Valseth}

\affiliation[inst1]{organization={The Department of Data Science, The Norwegian University of Life Science},
            addressline={Drøbakveien 31},
            city={Ås},
            postcode={1433},
            country={Norway}}
\affiliation[inst2]{organization={The Oden Institute for Computational Engineering and Sciences, The University of Texas at Austin},
            addressline={201 E. 24th St. Stop C0200},
            city={Austin},
            postcode={78712},
            state={Texas},
            country={United States of America}}
\affiliation[inst3]{organization={Coastal Resilience Center, University of North Carolina at Chapel Hill},
            city={Chapel Hill},
            postcode={27517},
            state={North Carolina},
            country={United States of America}}
\affiliation[inst4]{organization={Institute of Marine Sciences, University of North Carolina at Chapel Hill},
            city={Morehead City},
            postcode={28557},
            state={North Carolina},
            country={United States of America}}
\affiliation[inst5]{organization={The Department of Civil, Environmental, and Geodetic Engineering, The Ohio State University},
            addressline={2070 Neil Ave},
            city={Columbus},
            postcode={43210},
            state={Ohio},
            country={United States of America}}
\affiliation[inst6]{organization={The Department of Mechanical Engineering and Technology Management, The Norwegian University of Life Science},
            addressline={Drøbakveien 31},
            city={Ås},
            postcode={1433},
            country={Norway}}
\affiliation[inst7]{organization={Department of Scientific Computing and Numerical Analysis, Simula Research Laboratory},
            addressline={Kristian Augusts gate 23},
            city={Oslo},
            postcode={0164},
            country={Norway}}

\titlepage
\begin{abstract}

Recent severe storms on the U.S. Gulf coast have demonstrated the challenges presented by compound flooding, such as the interactions between rainfall runoff and storm surge. Historically, many studies have neglected these nonlinear interactions, but we propose to use a discontinuous Galerkin shallow water equation solver, which allows for incorporation of rainfall inputs directly onto the finite element mesh. In this work, we analyze the use of parametric rainfall for forecasting scenarios, using Hurricane Beryl (2024) as a case study. Beryl led to extensive flooding due to rainfall and storm surge along the Gulf. We use a collection of the National Oceanic and Atmospheric Administration's short-term advisories along with the best track data to demonstrate the efficacy of the parametric rainfall model for forecasting.

Results show that the parametric rainfall input allowed for much more accurate inundation. Areas with heavy rainfall and low surge were affected the most, with many areas peaking over 50 cm above the baseline surge model. 
Almost none of the available high water marks from Beryl were captured by the standard models, but the compound flooding models capture many of them, the majority of which show relative errors under 10 percent. Results from the advisory forecast simulations were shown to be much closer to the best track hindcast simulation when rainfall forcing was used, even while early forecasts predicted the storm's trajectory much less accurately. The advisory simulations improved even further as Beryl neared the Texas coast, with sampled peak elevations most closely approximating the best track at Advisory 38, a few hours before landfall.

\end{abstract}
\end{frontmatter}

\section{Introduction} \label{sec:introduction}

The Texas coast is characterized by a complex, low-lying system of barrier islands and shallow lagoons that separates the Gulf of Mexico from the continental mainland. From Matagorda Bay in the southwest to Galveston Bay in the northeast, the shoreline consists of long, sandy barriers, including the Matagorda Peninsula, South Padre Island, and Galveston Island. These barrier islands are also interrupted by natural and man made tidal inlets such as, Aransas Pass, San Luis Pass, and the Bolivar Roads \cite{Kennedy2011, Anderson2014}. Between these barriers and the mainland lie vast, shallow bay systems with average depths often less than 3 meters \cite{matlock1982shoreline}, extensive marshes and the dense urban topography, e.g., of the Greater Houston metropolitan area. The entire region sits atop a wide, gently sloping continental shelf that extends nearly 200 km offshore \cite{hope2013hindcast}. This bathymetric profile is particularly efficient at generating storm surge; wind stress acting over the broad shelf drives significant Ekman transport toward the coast, piling water against the barrier islands and forcing it through inlets into the bay systems where it can remain trapped for longer periods after a storm has passed  \cite{Rego2010, Sebastian2014}. 

Lying in a region of high tropical storm activity, the Texas coast region has been subject to storm driven coastal flooding throughout history~\cite{monica20244500}. Extreme historical floods include the Great Galveston Hurricane (1900), Hurricane Beulah (1968), Hurricane Ike (2008), and Hurricane Harvey (2017). The relatively flat topography of the region makes the Texas coast particularly vulnerable to storm surges. The Great Galveston Hurricane (1900), which remains the deadliest natural disaster in United States history, demonstrated the sheer lethality of surge when it inundated Galveston,  and completely covered the low-lying terrain \cite{Garrett2013}. Similarly, Hurricane Carla (1961) produced one of the largest surges on record for the Texas coast, of over 6 meters and reshaped the barrier islands on the coast \cite{Hayes1967}.
While remembered as a surge event, Hurricane Ike (2008)~\cite{ike2008}, was also characterized by a ``forerunner''~\cite{hope2013hindcast} wave that arrived days before landfall. This highlighted the challenges of the Texas bay system to drain efficiently once elevated \cite{Kennedy2011}.

Hurricane Harvey (2017)~\cite{harvey2017} initially made landfall near Port Aransas, Texas, and led to storm surge and wind damage in the southern part of the Texas Coast. However, the storm system eventually moved north and stalled near the Houston-Galveston area and resulted in extreme rainfall. In Galveston Bay, the minor surge prevented drainage of the inland runoff and induced catastrophic compound floods~\cite{ValleLevinson2020}. 
Since Hurricane Harvey (2017), other storms have also had a similar compound flood response, e.g.,  Tropical Storm Imelda (2019)~\cite{Latto2020}, and Tropical Storm Beta (2020)~\cite{Berg2021}. Further south on the coast, Tropical Storm Allison (2001) is another prominent compound flood example in Texas~\cite{OConnor2001}.
Most recently, Hurricane Beryl (2024)~\cite{nhc2024beryl} led to extensive flooding due to rainfall and thus fits into this emerging pattern of tropical storms with extensive rainfall and resulting floods.

Other well known compound flood events due to tropical storms in the last decade include Hurricane Irma (2017)~\cite{cangialosi2018tropical} and Hurricane Florence (2018)~\cite{callaghan2020extreme}. The flow resulting from the interaction of runoff and surge during such events processes requires careful treatment to ensure accurate modeling and has therefore been frequently studied in recent literature; see, e.g.,~\cite{loveland2021developing,santiago2019comprehensive,orton2020flood,kumbier2018investigating,wichitrnithed2024discontinuous}. Our goal in the current paper is to validate and comprehensively assess an existing numerical model in a compound flood forecasting of tropical cyclones~\cite{wahl2015increasing}.


Compound flooding is not limited to floods resulting from the interaction of surge and runoff; another potential source of floodwater is groundwater~\cite{Moftakhari2017}.
However, in this paper we limit our focus to compound floods resulting from storm surge and rainfall runoff.  
Despite the clear physical mechanisms of compound flooding in this setting, accurate real-time forecasting of compound events remains a significant computational challenge. Historically, operational frameworks have relied on ``loosely coupled'' approaches, where atmospheric, ocean, and hydrologic models run sequentially~\cite{Bakhtyar2020,Pena2022,loveland2021developing,pachev2023one}. While effective, this unidirectional data flow can fail to capture the dynamic synchronization of surge and runoff that defines events like Hurricanes Harvey (2017) or Beryl (2024)~\cite{Ye2020}.

In recent years, the modeling paradigm has shifted toward fully integrated modeling frameworks that solve the shallow water equations (SWE) for both coastal and overland flow within a single computational domain~\cite{santiago2019comprehensive}. Research indicates that neglecting the nonlinear interaction between surge and rainfall can underestimate flood extent by 20--30\% in transition zones \cite{Moftakhari2017, Gori2020}. To address this, there is a critical need for hydrodynamic solvers capable of internally generating realistic compound forcing from available inputs, thereby maintaining physical accuracy while eliminating the computational bottleneck of external coupling, e.g., as done by Wichitrnithed~\emph{et al.}~\cite{wichitrnithed2024discontinuous}.

To simulate the compound flooding due to rainfall and storm surge during a tropical storm, we rely on the discontinuous Galerkin SWE Model (DG-SWEM)~\cite{kubatko2006hp,bunya2009wetting,dawson2011discontinuous}, which solves the conservative SWE using a local discontinuous Galerkin (DG) finite element method with an explicit strong stability preserving Runge Kutta time stepping scheme. In the recent paper \cite{wichitrnithed2024discontinuous}, this model was extended to incorporate rainfall onto the FE mesh to model compound floods. Preliminary results indicated that DG-SWEM was capable of accurate hindcasting of compound floods. The key advancement in \cite{wichitrnithed2024discontinuous} was to modify the right-hand side of the continuity equation in the SWE with a source defined by parametric rainfall models thereby enabling on-the-fly generation of rainfall fields.

As Hurricane Beryl occurred in 2024, the peer-reviewed literature regarding hydrodynamic simulations of the event is currently in an emerging state. \cite{nabukulu2024} presented one of the first specific flood simulations, applying a spatial-temporal rainfall disaggregation tool to model pluvial hazard pathways in Grenada. Whereas, from an operational perspective, \cite{ccrif2024} utilized excess rainfall models to simulate wind-driven surge and rainfall accumulation across the Caribbean to calculate insurance payout triggers. On the Texas coast, the technical report \cite{hcfcd2024beryl} provides the essential hydrologic boundary conditions for future compound flood modeling, and includes documentation for the blocking mechanism where surge elevations in the Houston Ship Channel prevented the drainage of the runoff.

In this paper, we present the investigation of the use of DG-SWEM in a forecasting scenario with Hurricane Beryl (2024) as a case study. In Section~\ref{sec:beryl}, we describe Hurricane Beryl and its timeline. In Section~\ref{sec:swe}, we describe DG-SWEM and the approach taken to forecasting. The results from our simulation campaign of Hurricane Beryl (2024) are presented in Section~\ref{sec:Numerical_results}. Finally, conclusions and comments on future works are in Section~\ref{sec:conclusions}.

\section{Hurricane Beryl (2024) and Study Area} \label{sec:beryl}

\begin{figure}
    \centering
    \includegraphics[width=0.8\linewidth]{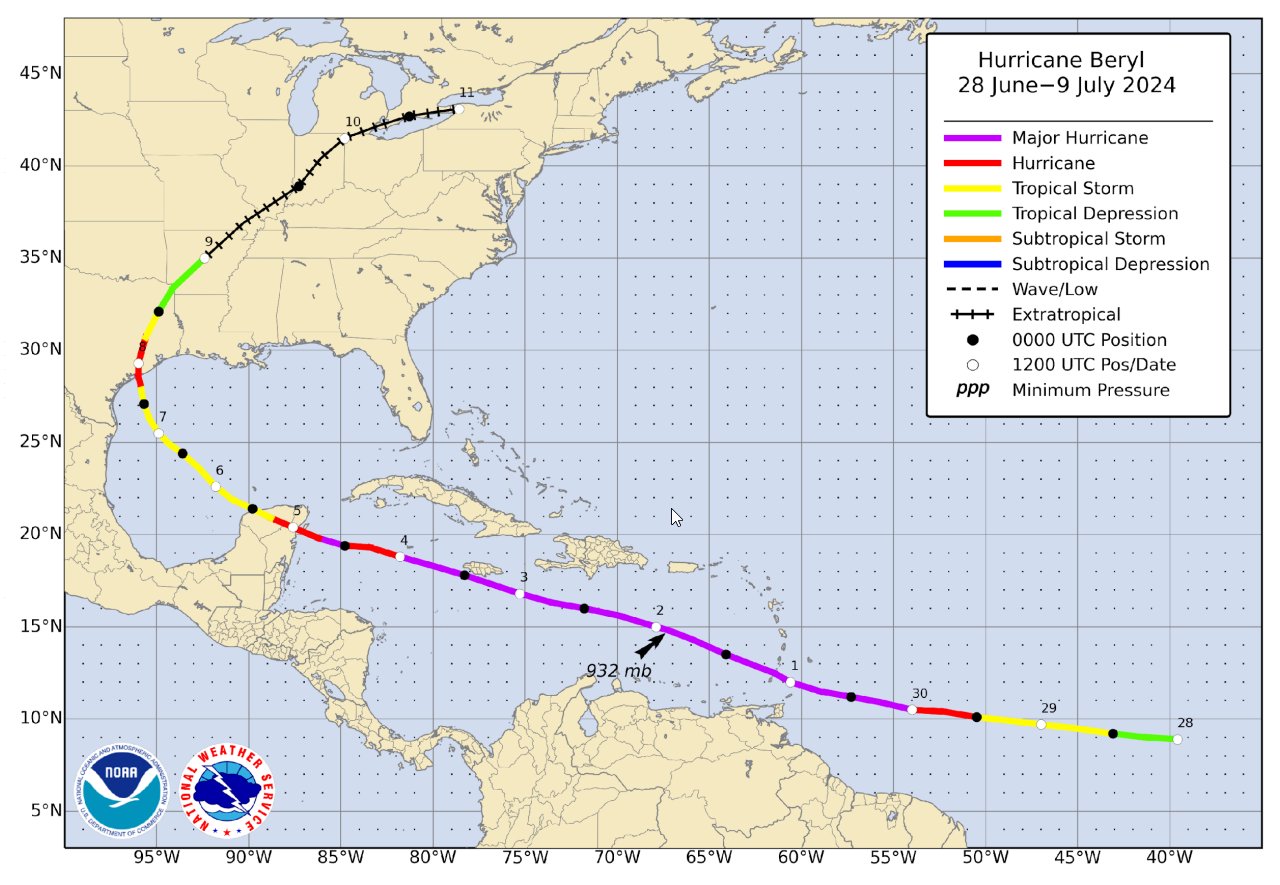}
    \caption{Best track data for Hurricane Beryl, from the NHC \cite{nhc2024beryl}}
    \label{fig:track}
\end{figure}

Hurricane Beryl originated from a tropical wave that departed the west coast of Africa on June 25, 2024. The system intensified into a tropical depression on June 28. In a record-breaking phase of rapid intensification, and became the earliest Category 5 hurricane on record in the Atlantic basin, reaching a peak intensity of 266 km/h and a minimum central pressure of 932 mbar on July 2 while in the eastern Caribbean Sea \cite{nhc2024beryl}. The best track positions from the National Hurricane Center (NHC) are shown in Figure \ref{fig:track}.

After impacting Caribbean islands and the Yucatán Peninsula, Beryl moved northwestward across the Gulf toward the Texas coast. The storm regained Category 1 hurricane status just hours before landfall near Matagorda, Texas, at approximately 08:40 UTC on July 8, 2024, with maximum sustained winds of 148 km/h and a central pressure of 979 mbar.
Following landfall, the storm center tracked north-northeastward, passing directly over the western Houston metropolitan area. While the wind field decayed inland, the system maintained a robust circulation that facilitated the transport of deep tropical moisture from the Gulf, resulting in heavy rainfall that led to the compound nature of Hurricane Beryl~\cite{nhc2024beryl}. The timeline of the storm's intensity and position is summarized in Table~\ref{tab:beryl_track}.

\begin{table}[ht]
    \centering
    \caption{Best track positions and intensity for Hurricane Beryl (2024). Data sourced from the NHC Tropical Cyclone Report \cite{nhc2024beryl}.}
    \resizebox{\textwidth}{!}{
    \begin{tabular}{l c c c c l} 
        \toprule
        \textbf{Date / Time (UTC)} & \textbf{Lat ($^\circ$N)} & \textbf{Lon ($^\circ$W)} & \textbf{Press. (mb)} & \textbf{Wind (kt)} & \textbf{Status} \\
        \midrule
        28 Jun / 12:00 & 9.4 & 42.6 & 1006 & 30 & Tropical Depression \\
        29 Jun / 18:00 & 10.0 & 49.3 & 992 & 65 & Hurricane (Cat 1) \\
        01 Jul / 15:20 & 12.5 & 61.5 & 950 & 130 & Landfall (Carriacou) \\
        02 Jul / 06:00 & 13.8 & 65.2 & 932 & 145 & Peak Intensity (Cat 5) \\
        04 Jul / 12:00 & 18.5 & 80.5 & 962 & 110 & Passing Cayman Is. \\
        05 Jul / 11:05 & 20.2 & 87.4 & 975 & 75 & Landfall (Yucatán) \\
        06 Jul / 12:00 & 22.0 & 91.0 & 997 & 50 & Tropical Storm (Gulf) \\
        08 Jul / 00:00 & 26.8 & 95.3 & 986 & 60 & Re-intensifying \\
        08 Jul / 04:00 & 27.8 & 95.7 & 982 & 65 & Hurricane (Cat 1) \\
        \textbf{08 Jul / 08:40} & \textbf{28.6} & \textbf{96.0} & \textbf{979} & \textbf{80} & \textbf{Landfall (TX)} \\
        08 Jul / 12:00 & 29.3 & 96.0 & 985 & 65 & Inland (Wharton) \\
        09 Jul / 00:00 & 32.1 & 94.7 & 996 & 35 & Tropical Depression \\
        \bottomrule
    \end{tabular} }
    \label{tab:beryl_track}
\end{table}

\subsection{Flood Impacts on the Texas Coast}

A key driver of the compound flooding during Hurricane Beryl was the  wind field as it interacted with the coastline. Beryl made landfall near Matagorda, Texas, as a Category 1 hurricane with sustained winds of approximately 129 km/h and gusts reaching 172 km/h \cite{nhc2024beryl, nws2024lch}. Unlike broader, weaker systems, Beryl was compact and undergoing rapid intensification at the moment of landfall~\cite{nhc2024beryl}. The storm maintained hurricane-force wind gusts well inland; e.g., gusts of 134 km/h were recorded as far north as George Bush Intercontinental Airport in Houston~\cite{nws2024hgx}. This sustained wind stress over land was critical in driving wind-forced surge into narrow inland channels like the Houston Ship Channel~\cite{hcfcd2024beryl}.

Despite being classified a Category 1 storm at landfall, Hurricane Beryl produced significant surge levels, ranking as the highest storm surge in the region since Hurricane Ike (2008) \cite{hcfcd2024beryl}. The surge reached a record-breaking 1.66 meters above Mean Higher High Water (MHHW) at Morgans Point \cite{noaa2024morgans}, with levels reaching nearly 2.7 meters above ground level at the Ship Channel Turning Basin \cite{hcfcd2024beryl}. These high water levels were driven by the storm's track, which allowed the winds to force water directly into Galveston Bay~\cite{hcfcd2024beryl}. While the open coast from Matagorda to Freeport experienced moderate surge levels, the peak water levels were amplified within the confined bay system due to this funnel effect~\cite{nws2024lch}.

A defining characteristic of Beryl's impact was the compound flooding due to storm surge and intense rainfall. Beryl deposited 15--36 centimeters of rain over the Houston metropolitan area, with a maximum 24-hour total of 30 centimeters recorded at Houston Transtar~\cite{hcfcd2024beryl}. Under normal conditions, this runoff drains into major bayous into Galveston Bay. However, the surge in the Ship Channel acted as a blocked this drainage and elevated the water levels in these drainage systems~\cite{hcfcd2024beryl}. 

The coastal flooding was further aggravated by the sea state and the condition of the coastline prior to landfall. The U.S. Geological Survey projected that 68\% of the Texas coastline would experience erosion due to the storm \cite{usgs2024erosion}. The coast was primed by Tropical Storm Alberto, which had impacted the region just 18 days prior, eroding dunes and reducing the natural barriers available to mitigate flooding during Beryl~\cite{usgs2024erosion}.

\section{Numerical Model and Proposed Forecasting Approach } \label{sec:swe}

The model used in this work is based on the governing two-dimensional SWE, which consist of the depth-averaged equations of mass conservation as well as $x$ and $y$ momentum conservation \citep{tan1992shallow}. The SWE are stated in conservative form in Equation \eqref{eq:SWE}.

\begin{figure}
    \centering
    \begin{tikzpicture}[scale=1.4]

    \definecolor{waterblue}{RGB}{156,195,222}
    \definecolor{sand}{RGB}{148,147,108}
    
    \draw[black, very thick] (-4,0) -- (4,0);
    
    \draw[brown, thick] plot[smooth, tension=.7] coordinates {(-4,-2.5) (-2,-1.8) (0,-2.3) (2,-2.0) (4,-2.4)};
    \fill[sand] plot[smooth, tension=.7] coordinates {(-4,-3) (-4,-2.5) (-2,-1.8) (0,-2.3) (2,-2.0) (4,-2.4) (4,-3)} -- cycle;

    \draw[blue, thick] plot[smooth, tension=.7] coordinates {(-4,1.0) (-2,0.8) (0,1.2) (2,1.1) (4,1.3)};
    \fill[waterblue] plot[smooth, tension=.7] coordinates {(-4,1.0) (-2,0.8) (0,1.2) (2,1.1) (4,1.3)} -- plot[smooth, tension=.7] coordinates { (4,-2.4) (2,-2.0) (0,-2.3) (-2,-1.8) (-4,-2.5) } -- cycle;
    
    \node[right] at (1, -0.6) {$\qquad H = \zeta + h_b$};
    
    \node[right] at (4.5, 0.450) {$\quad \zeta $};
    \node[right] at (4.5, -1.2) {$\quad h_b$};

    \draw[<->] (1.3, 1.1) -- (1.3, -2.1);

    \draw[<->] (4.5, 1.3) -- (4.5, 0);
    \node[above] at (4.5, 1.3) { };

    \draw[<->] (4.5, 0) -- (4.5, -2.4);
    \node[below] at (4.5, -2.4) { };
    
    \draw[black , thick] (-4,0) -- (4,0);

\end{tikzpicture}
    \caption{Definition of shallow water elevations. The horizontal line is the geoid, where $\zeta = h_b = 0$.}
    \label{fig:elevation_def}
\end{figure}
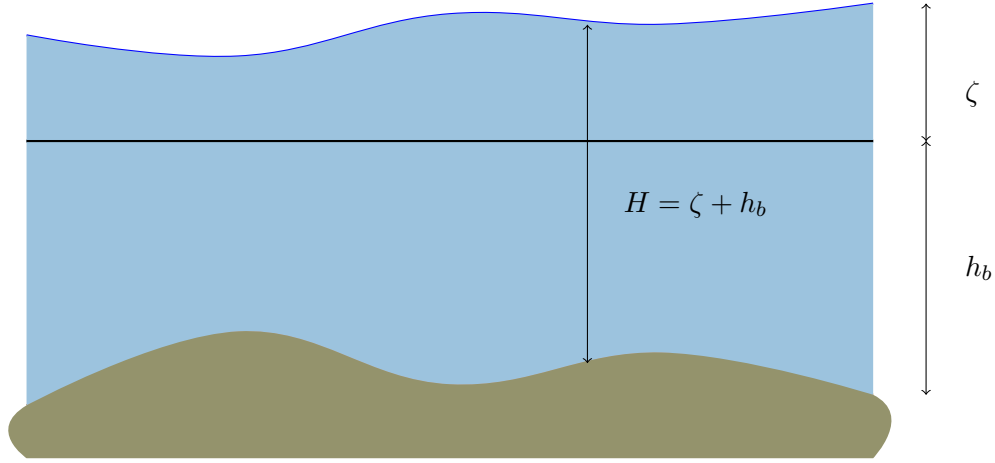

{Find }  $(\zeta, \mathbf{u})$  { such that:}  
\begin{align} 
\begin{split}
    \displaystyle \frac{\partial  \zeta}{\partial t} + \nabla \cdot (H{\mathbf{u}})  &= R, \text{ in } \Omega, \\
    \displaystyle\frac{\partial (Hu_x)}{\partial t} + \nabla \cdot \left( Hu_x^2 + \frac{g}{2}(H^2-h_b^2), Hu_xu_y \right) - g\zeta \frac{\partial h_b}{\partial x} + \kappa u_x &= F_x, \text{ in } \Omega, \\ 
    \displaystyle\frac{\partial (Hu_y)}{\partial t} + \nabla \cdot \left( Hu_xu_y, Hu_y^2 + \frac{g}{2}(H^2-h_b^2) \right) - g\zeta \frac{\partial h_b}{\partial y} + \kappa u_y &= F_y, \text{ in } \Omega,
\end{split} \tag{1} \label{eq:SWE}
\end{align}
\setcounter{equation}{1}
where $\zeta$ represents the free surface elevation (positive upwards from the geoid), $h_b$ is the bathymetry (positive downwards from the geoid), and $H$ is the total water column, as illustrated in Figure \ref{fig:elevation_def}. $\mathbf{u} = \{ u_x,u_y\}^{\text{T}}$ represents the depth-averaged horizontal velocity field, $\kappa$ is the bottom friction factor, and $R$ is the continuity equation source term, which represents rainfall in our model. $F_x$ and $F_y$ represent other forces, which include tidal potential forces, wind stresses, wave radiation stresses, vertically integrated lateral stresses, and Coriolis forces ~\citep{luettich2004formulation}.
The bottom friction terms can assume a linear or nonlinear form, or a mix of both; in this study, we use Manning's formula for bottom friction \cite{Manning1891}.  
The computational domain is denoted by $\Omega$, and its boundary $\Gamma$ is specified by two distinctive sections $\Gamma = \Gamma_{ocean}\cup\Gamma_{land}$. On these sections, the boundary conditions include specified (tidal) elevation conditions and zero normal flow, respectively. Other applicable boundary conditions include specified normal flow, i.e, river flow, though in the present work we do not consider other boundary conditions.

These partial differential equations are used to model flows in the coastal regions, estuaries, and rivers, where the vertical dimension is much smaller than the horizontal scale of the flow. 

\subsection{Discretization}

The numerical approximation of the SWE is accomplished utilizing the discontinuous Galerkin method, and we describe that discretization briefly here (c.f. \cite{kubatko2006hp, wichitrnithed2024discontinuous}). Lumping bottom friction forces into $F_{x}$ and $F_{y}$, we write equation \ref{eq:SWE} in divergence form as follows:
\begin{align}
    \frac{\partial }{\partial t} \underbrace{\begin{bmatrix}
        \zeta \\
        u_{x}H \\
        u_{y}H
    \end{bmatrix}}_{\mathbf{w}} + \nabla \cdot \underbrace{\begin{bmatrix}
        Hu_{x} & Hu_{y} \\
        Hu_{x}^{2} + g \left( H^{2} - h_{b}^{2} \right) & Hu_{x}u_{y} \\
        Hu_{x}u_{y} & Hu_{y}^{2} + g \left( H^{2} - h_{b}^{2} \right)
    \end{bmatrix}}_{\mathbf{F}(\mathbf{w})} &= \underbrace{\begin{bmatrix}
        R \\
        g \zeta \frac{\partial h_{b}}{\partial x} + F_{x} \\
        g \zeta \frac{\partial h_{b}}{\partial y} + F_{y}
    \end{bmatrix}}_{\mathbf{s}} \label{eq:div}
\end{align}
The vector of unknowns, $\mathbf{w}$, is approximated by $\mathbf{w}_{h}$, which is defined as piecewise differentiable over each element, but allows for discontinuities between elements; we denote this function space as $V_{h}$. We obtain the discrete weak form of the equation by replacing $\mathbf{w}$ with $\mathbf{w}_{h}$, multiplying by a test function $\mathbf{v} \in V_{h}$, integrating over each element $e$, then integrating by parts:
\begin{align}
    \left( \frac{\partial}{\partial t} \left( \mathbf{w}_{h}^{(i)} \right), v \right)_{\Omega_{e}} - \left( \nabla v, \mathbf{F}_{i} \right)_{\Omega_{e}} + \langle \hat{\mathbf{F}}_{i} \cdot \hat{\mathbf{n}}, v \rangle_{\Gamma_{e}} &= \left(s, v \right)_{\Omega_{e}} \label{eq:weakform}
\end{align}
where $\mathbf{w}_{h}^{(i)}$ is the $i$th component of $\mathbf{w}_{h}$, $\hat{\mathbf{n}}$ is the unit normal vector of $\Gamma$, and numerical flux $\hat{\mathbf{F}}$ is used in place of $\mathbf{F}$ in the boundary integral, since $\mathbf{F}$ may be dual valued along $\Gamma$. Several choices of numerical flux are available, and this study employs Local Lax-Friedrichs flux, which is shown to be both stable and accurate \cite{Li2020-llf}. The piecewise differentiable $\mathbf{w}_{h}$ is approximated using an orthogonal Dubiner basis, as
\begin{align}
    \mathbf{w}_{h} &= \sum_{i} \sum_{j} \tilde{\mathbf{w}}_{ij} \phi_{ij},
\end{align}
where $\tilde{\mathbf{w}}_{ij}$ are the modal degrees of freedom and $\phi_{ij}$ are polynomials of order $i + j$. This choice of basis results in a diagonal mass matrix, and allows for higher-order approximations to be obtained by adding additional modes to the basis functions $\phi$.

Substituting $\mathbf{w}_{h}$ into Equation \ref{eq:weakform} yields the reduced problem, in the form of a system of ordinary differential equations in time, i.e.:
\begin{align}
    \frac{\mathrm{d}}{\mathrm{d}t} \left(\mathbf{w}_{h} \right) &= L_{h} \left( \mathbf{w}_{h} \right).
\end{align}

The time discretization for this system is based on a total variation diminishing Runge--Kutta scheme \cite{Shu1988-eq}. The new Strong Stability-Preserving Runge-Kutta timestepping scheme optimizes the allowable timestep size, while maintaining high-order convergence and stability when applied to DG spatial discretizations \cite{Kubatko2014-nc}.

\subsection{Forecasting with Rainfall} \label{sec:rain_models}

Incorporating rainfall into the model introduces a few challenges that must be addressed. First, the rainfall does not introduce instability to the problem, since DG-SWEM uses the primitive continuity equation, and rainfall simply forms the source term $R$ \cite{wichitrnithed2024discontinuous}. Another key element to consider when adding rainfall directly to the mesh involves the wetting and drying algorithm. Rainfall inputs should work on both wet and dry elements. Because rainfall is added to the mesh before the wet/dry check is performed, this requirement is satisfied.

For forecasting purposes, we need to use predicted rainfall data from atmospheric models. During tropical cyclones, rainfall patterns exhibit characteristic structures that may be exploited by parametric rainfall models. The basic idea behind this type of approach is to use simple analytic expressions based on storm parameters from the NHC to construct a rainfall field. This type of approach has been implemented in DG-SWEM for a few models, including R-CLIPER (Rainfall CLImatology and PERsistence) \cite{Tuleya:2007} and IPET (Interagency Performance Evaluation Task Force Rainfall Analysis) \cite{ipet}, which compute rainfall rates $R(r)$ at any point in the grid, where $r$ is the distance from that point to the storm center. 

For the purposes of this paper, the R-CLIPER rainfall model is used. The rainfall rate is calculated as:
\begin{align}
    R(r) &= T_{0} + \left(T_{m} - T_{0}\right) \frac{r}{r_{m}} & \quad & r \leq r_{m} \\
    R(r) &= T_{m} e^{-\frac{r - r_{m}}{r_{e}}} & \quad & r > r_{m}
\end{align}
where $r_{m}$ is the radius of maximum rainfall, $r_{e}$ is the radius of the outer extent of the storm, $T_{0}$ is the maximum precipitation rate at the center of the storm, and $T_{m}$ is maximum precipitation rate, at radius $r_{m}$. These values are calculated by assuming a linear relationship between each parameter and a normalized maximum wind speed, then performing a least squares fit with data from a large collection of global tropical cyclones \cite{Tuleya:2007}.

\begin{figure}
    \centering
    \begin{subfigure}{0.45\textwidth}
        \centering
        \includegraphics[width=\textwidth]{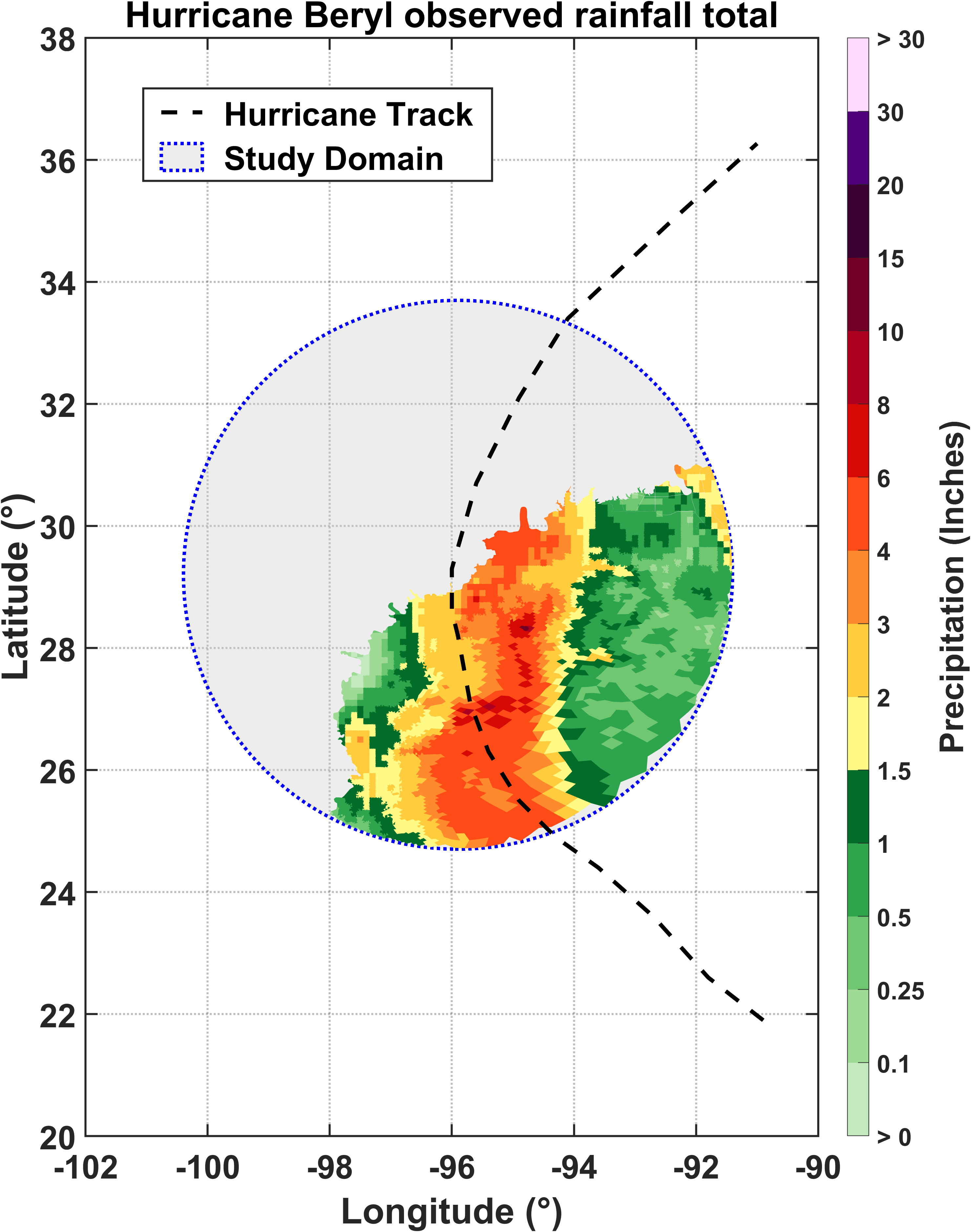}
    \end{subfigure}
    \begin{subfigure}{0.45\textwidth}
        \centering
        \includegraphics[width=\textwidth]{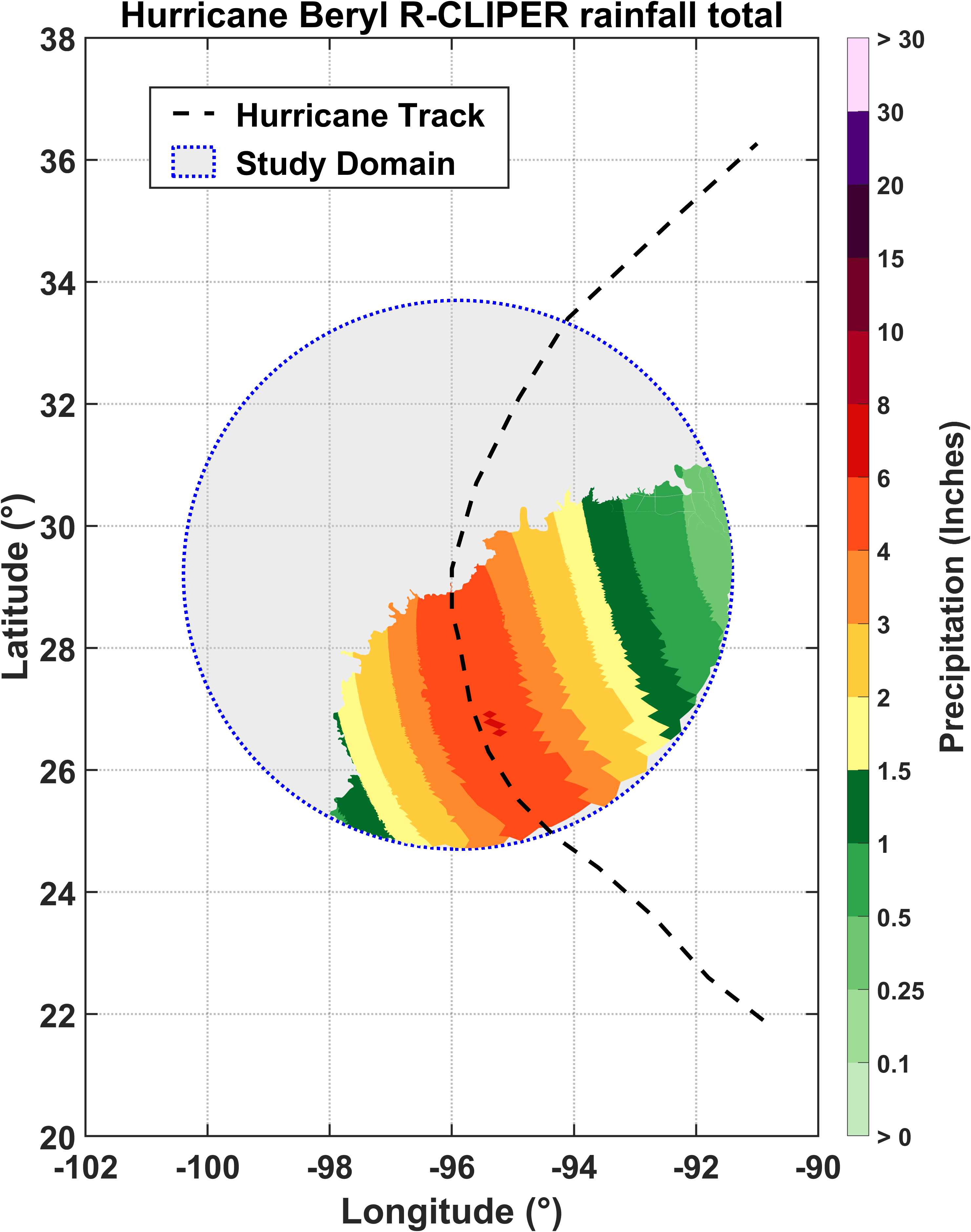}
    \end{subfigure}
    \caption{Example snapshot of R-CLIPER rainfall model during Hurricane Beryl}
    \label{fig:rcliper}
\end{figure}

The radially-symmetric R-CLIPER datapath is illustrated in Figure \ref{fig:rcliper}; see \citep{Tuleya:2007,Brackins2020-xp} for more details on general parametric schemes. The primary inputs of the model, the reported storm centers and maximum wind velocities, are generally provided at 6-hour intervals by the NHC, and thus could be used in real-time for forecasting. Within DG-SWEM, linear interpolation is used to obtain the storm parameters at the model time step, and the rainfall rate at each finite element mesh node is computed based on its radial distance from the storm center.
This rainfall rate is then transformed to the elemental source term by computing the average rain intensity over each of an element's three nodes:
\begin{align}
    R &= \frac{R \left( r (n_{e}^{1} \right) ) + R \left( r (n_{e}^{2} ) \right) + R \left( r (n_{e}^{3} ) \right)}{3} & \text{in } \Omega_{e}
\end{align}
where $r(n_{e}^{i})$ are the distances from the nodes of element $e$ to the storm center.

To analyze the use of this compound flooding simulation pipeline for forecasting, we run simulations for a number of different forecast advisories. Forty-two advisories were issued by the NHC, starting on June 28, 2024. Each advisory contains 3--5 days of wind and pressure forecasts.

DG-SWEM uses a modified version of the unstructured computational mesh developed in \cite{Contreras2023-qi}, called the East Coast Gulf of Mexico Coast Model (ECGC). This mesh incorporates many channels and a significant portion of low-lying land near the Texas coast, making it well-suited for compound flooding studies. The mesh was modified to be much less refined on the U.S. east coast, and more refined in the Gulf of Mexico. This results in a much smaller, more specific mesh containing 1,947,485 nodes, with 120-meter minimum resolution. The simulations are run on 2000 compute processors on the Frontera supercompter at the Texas Advanced Computing Center, with timestep size between 0.25 and 1 second. They all begin with an 8-day tidal spinup, before the atmospheric and rainfall forcing terms are added for their three- or five-day duration.
\section{Results} \label{sec:Numerical_results}

To examine the effect of rainfall on flood forecasting and to provide a baseline result, we first validate the model by simulating the best track hindcast both with and without rain. Then we may use existing gauge data, high water marks, and the best track hindcast model to analyze each of seven selected weather advisories, including two earlier ones a day apart (28 and 32), four advisories leading up to Beryl's landfall in Texas (36--39) and one immediately after landfall (40). 

\subsection{Validation} \label{sec:validation}

The full duration of Beryl was modeled with the best track wind data that spans a two-week period from 29 June to 11 July, with a timestep of 1 second. This best track hindcast, as well as the NOAA's own predictions and observed gauge data, form a diverse basis for validation and analysis.

Each forecast scenario was modeled both with and without the inclusion of the rainfall forcing term $R$, so that the effects of the rainfall may be measured. 

Results from the best track hindcasts are presented in Table \ref{tab:best_track_v_gauges}. When rainfall is added, we observe higher correlation between the model and the gauge data, as well as slightly decreased root mean square error (RMSE) on average.
\begin{table}
    \centering
    \small
    \caption{RMSE and correlation between the best track hindcasts and gauge data in the Houston area.}
    \begin{tabular}{c|c|c|c|c}
        \toprule
        & \multicolumn{2}{l|}{\textbf{RMSE} (m)} & \multicolumn{2}{l}{\textbf{Correlation Coefficient}}  \\
        \textbf{Elevation gauge} & With Rainfall & Without Rainfall & With Rainfall & Without Rainfall \\
        \midrule
        Galveston Railroad Bridge & 0.1745 & 0.1762 & 0.7969 & 0.7918 \\
        San Luis Pass & 0.2043 & 0.2047 & 0.7586 & 0.7556 \\
        Manchester Houston & 0.3952 & 0.3996 & 0.7483 & 0.7467 \\
        Rollover Pass & 0.2374 & 0.2370 & 0.6989 & 0.6970 \\
        Port Arthur & 0.1420 & 0.1417 & 0.8086 & 0.8028 \\
        Galveston Bay Entrance & 0.2048 & 0.2053 & 0.7905 & 0.7894 \\
        \bottomrule
    \end{tabular} 
    \label{tab:best_track_v_gauges}
\end{table}

\begin{figure}
    \centering
    \includegraphics[width=0.8\linewidth]{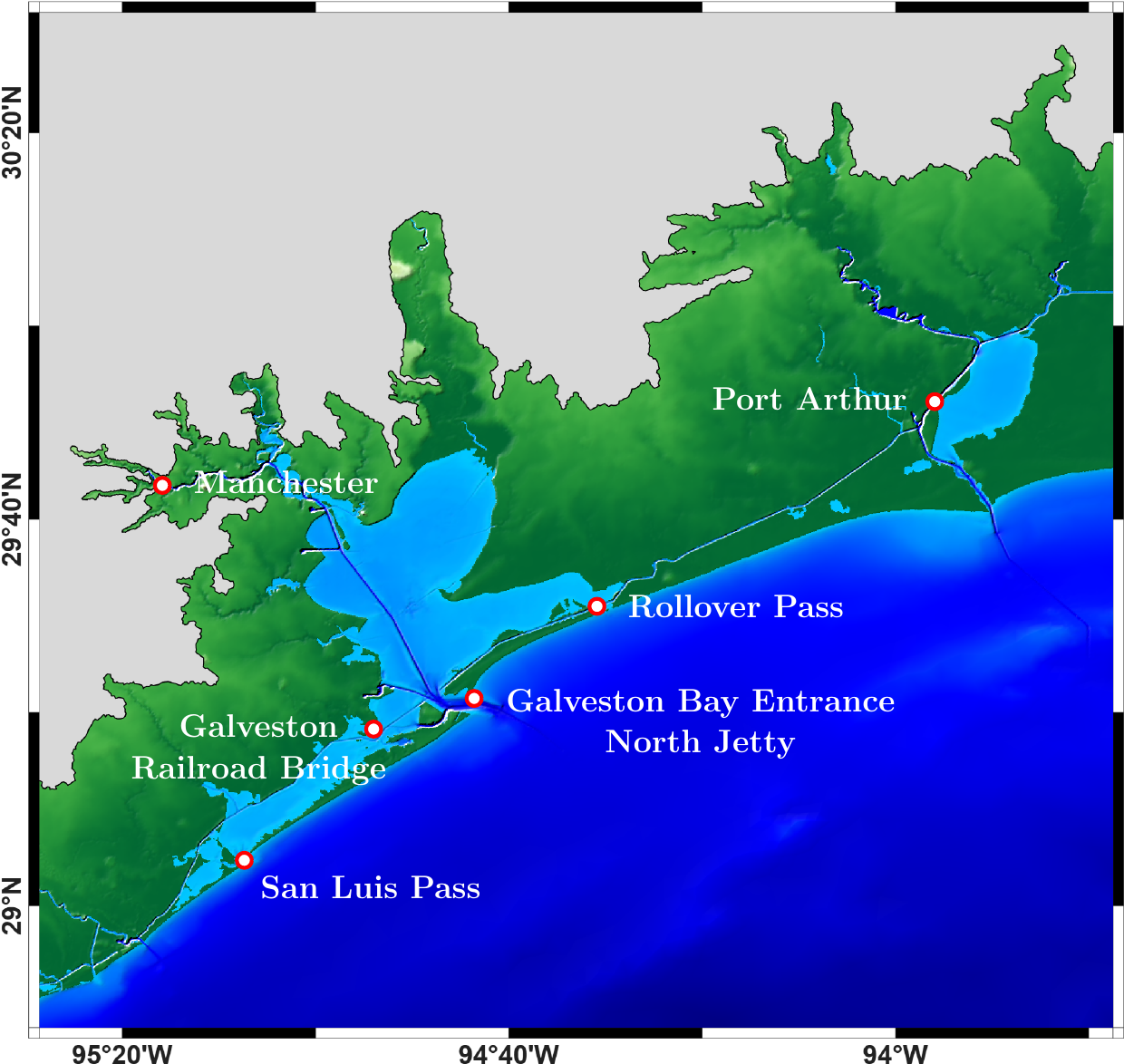}
    \caption{NOAA water surface elevation locations}
    \label{fig:gauges}
\end{figure}

We can compare each simulated advisory with water surface elevation gauge data. The gauges used are shown in Figure \ref{fig:gauges}.

\begin{figure}
    \centering
    \begin{subfigure}{0.45\textwidth}
        \centering
        \includegraphics[width=\textwidth]{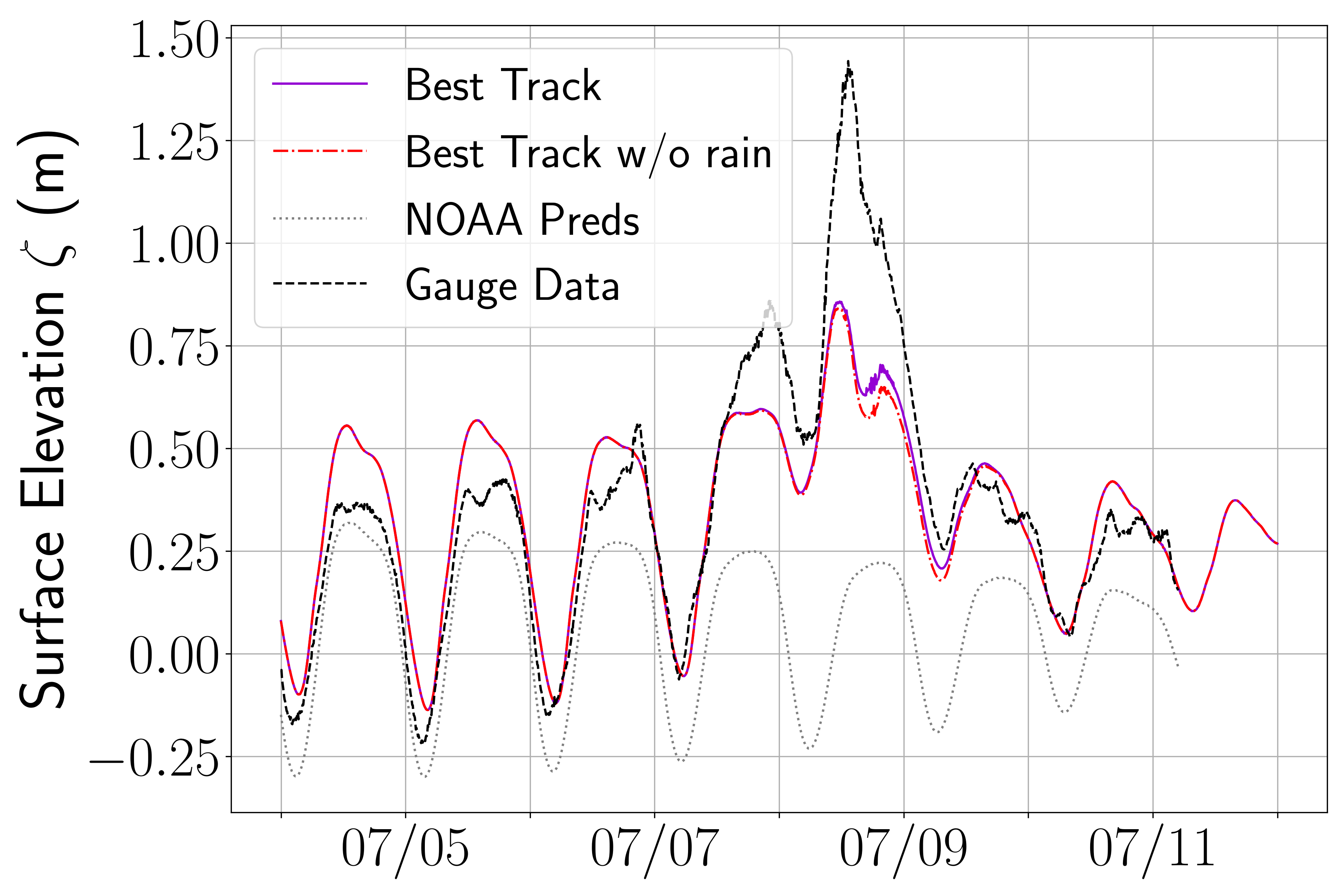}
        \caption{Water surface elevation tracks compared with gauge data}
    \end{subfigure}
    \begin{subfigure}{0.45\textwidth}
        \centering
        \includegraphics[width=\textwidth]{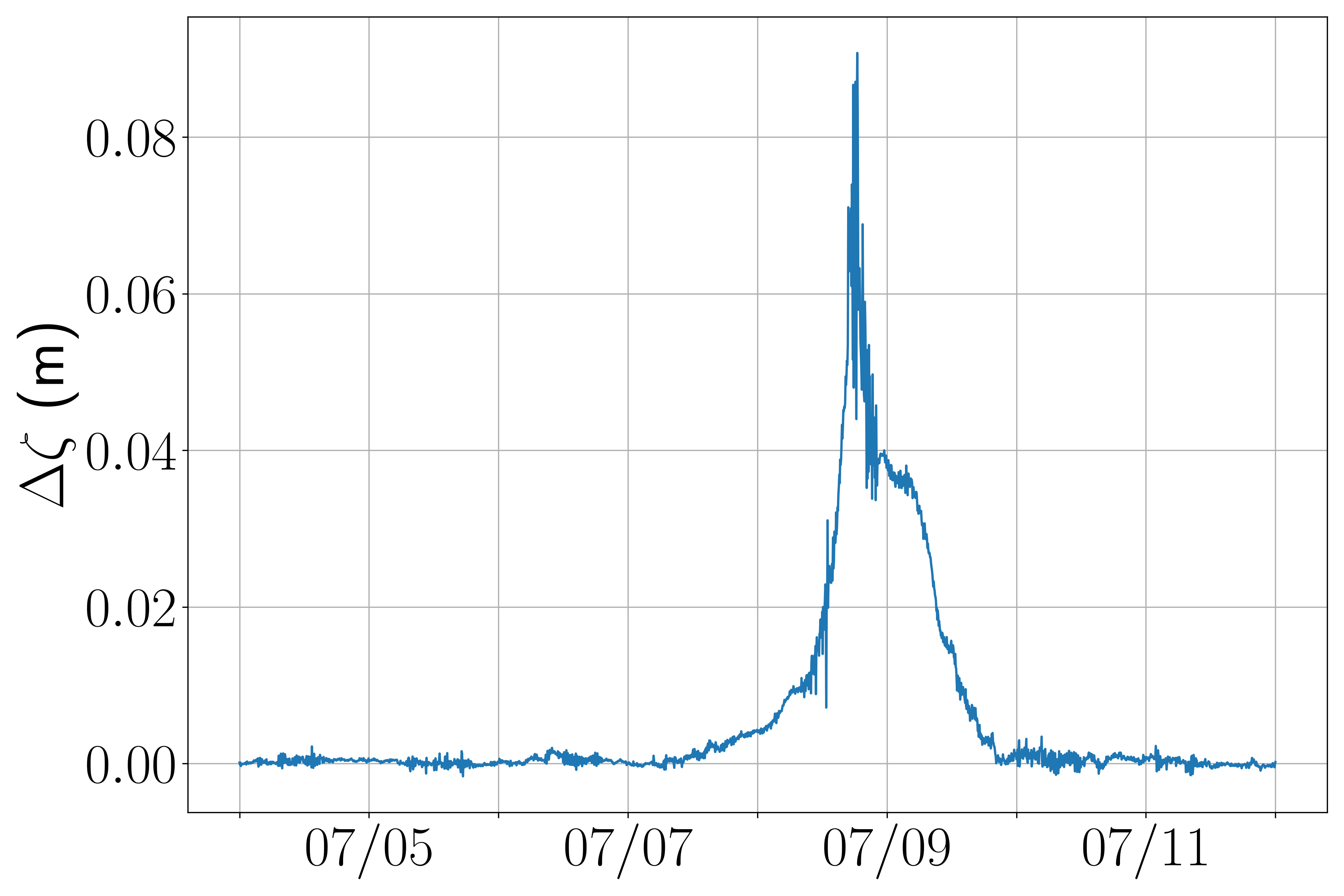}
        \caption{Difference between compound flooding and traditional models}
        \label{fig:galvestonrailroad_diff}
    \end{subfigure}    
    \caption{Hindcast results with and without rain, at the Galveston Railroad Bridge gauge}
    \label{fig:galvestonrailroad}
\end{figure}

For illustration, the best track hindcasts with and without rainfall forcing are shown together at the Galveston Railroad Bridge gauge, just inside Galveston Bay, in Figure \ref{fig:galvestonrailroad}. We observe that the two DG-SWEM models are practically identical until the extreme winds and rain from Beryl hit, and then a sustained offset appears for several hours, while the rain and its impact abates. The difference plot in Figure \ref{fig:galvestonrailroad_diff} reveals a spike on July 9, with a maximum difference of about 10 centimeters.

While the rainfall input has a clear effect throughout the domain, it is the most influential in upstream areas that are not reached by the storm surge the traditional model. Points with higher elevation above sea level are able to become wet due to the heavy rainfall added, and at the peak of the hurricane, there is an interaction between the surge and these formerly-wet nodes.

\begin{figure}
    \centering
    \includegraphics[width=0.8\textwidth]{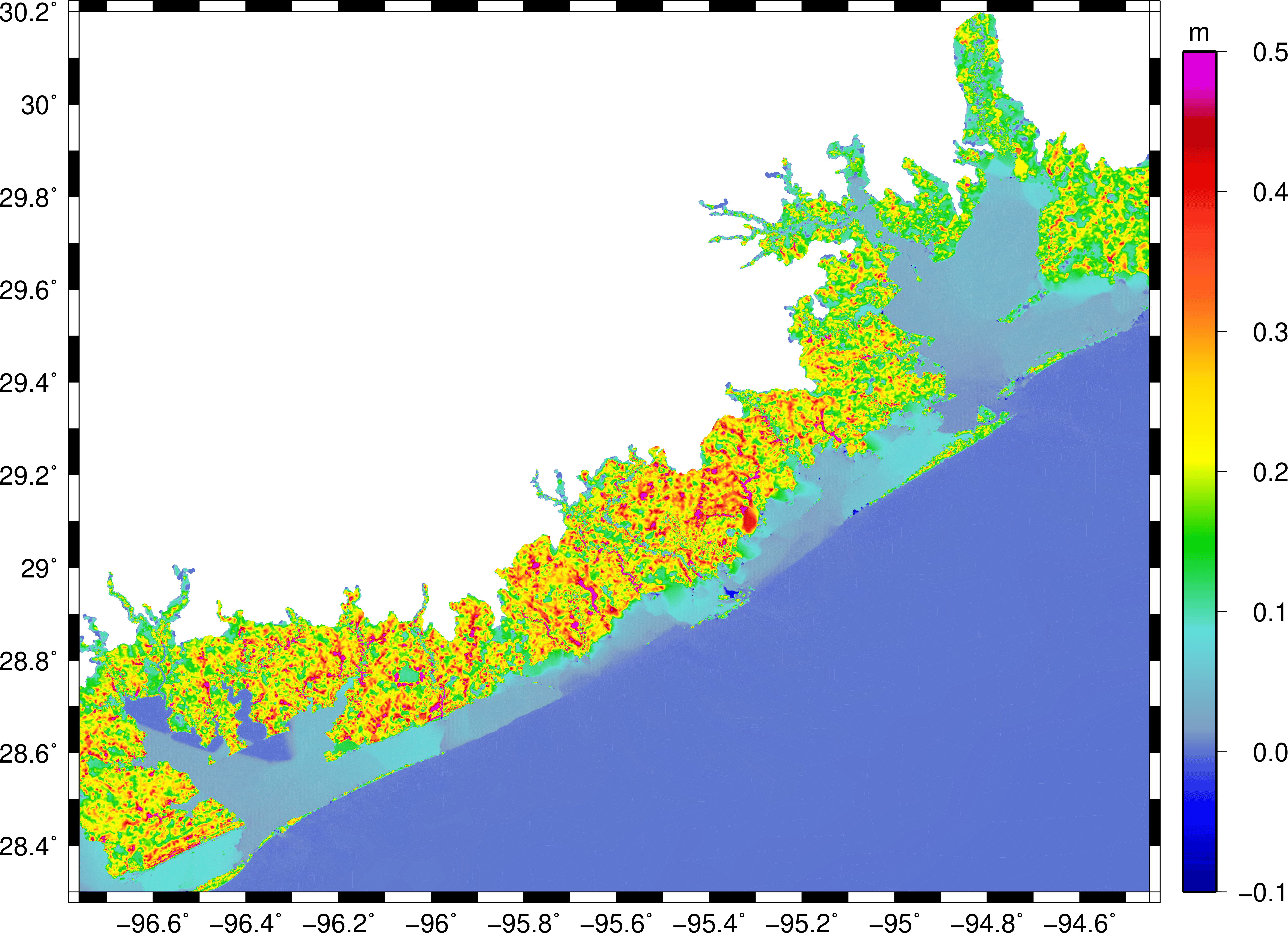}
    \caption{Difference in water column resulting from parametric rainfall input, between Matagorda and Galveston Bays}
    \label{fig:diff_map}
\end{figure}

Since the computational mesh contains many inland points several meters above sea level, simply taking a difference between the water surface elevations with and without rainfall forcing would be misleading. Instead, when a node in either model is dry, we set its maximum elevation to the node's elevation above sea level. Then we can show the additional water column above land as a result of the rainfall forcing. Figure \ref{fig:diff_map} shows this difference over the entire best track simulations.

We observe that the rainfall input resulted in tens of centimeters of additional water in low-lying areas along the coast, including several large regions of 40--50 centimeters of additional water. Notable increases are seen in the marshlands between Freeport, and Palacios, Texas, between the watersheds of the Brazos, San Bernard, Colorado, and Tres Palacios Rivers. Over three hundred nodes in this region peak at above one meter of difference. Incorporating this same rainfall data into forecasting models could help predict this significant inundation where the surge-only model could not, allowing better protective measures to be taken.

\begin{figure}
    \centering
    \includegraphics[width=0.7\linewidth]{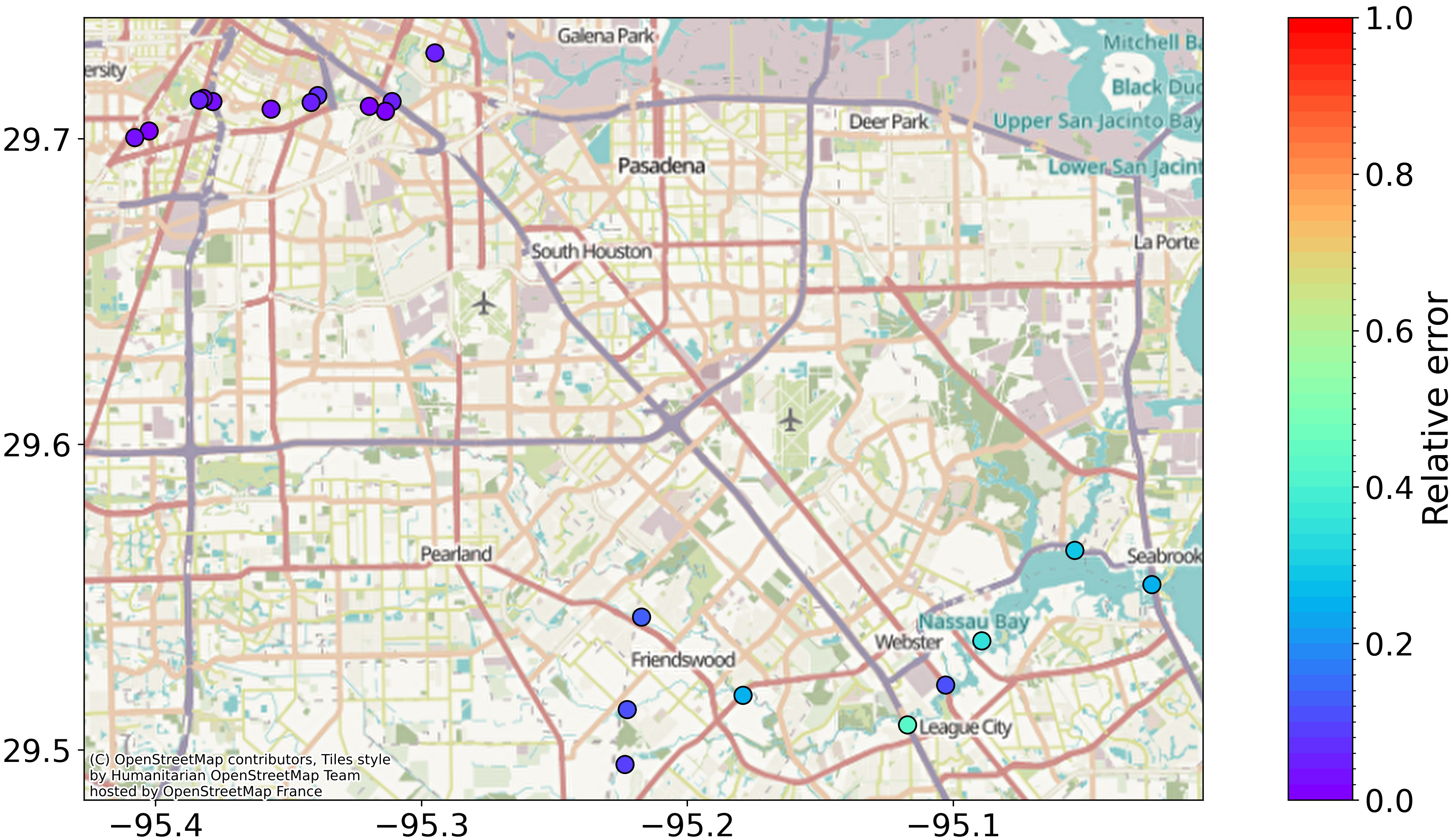}
    \caption{Accuracy of 25 high water marks from best track DG-SWEM hindcast with parametric rainfall. Map data from \cite{openstreetmap}} 
    \label{fig:hwm_rain}
\end{figure}

\begin{figure}
    \centering
    \includegraphics[width=0.7\linewidth]{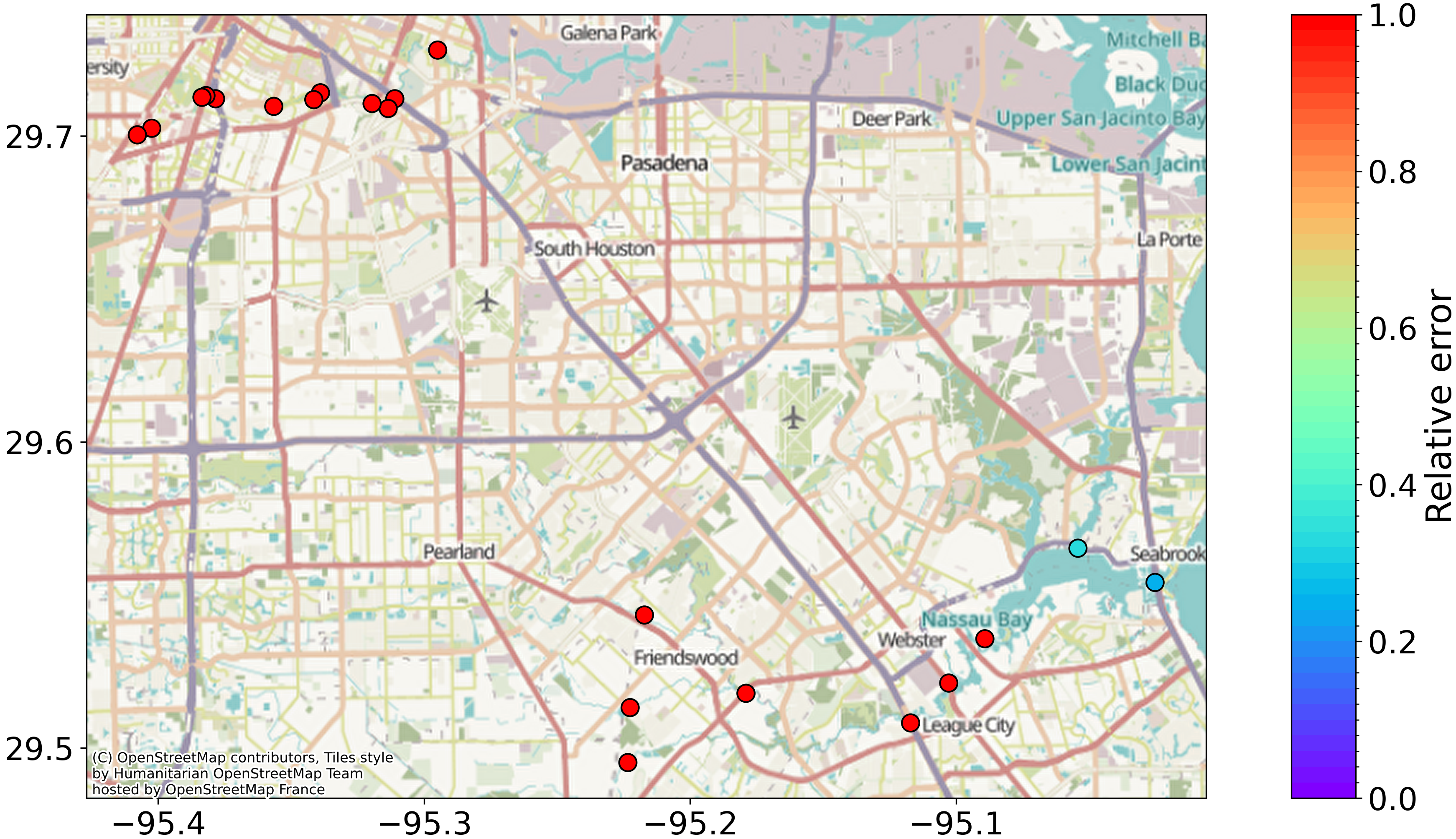}
    \caption{Accuracy of high water marks from best track DG-SWEM hindcast without rainfall forcing. Only two HWMs are wet in this model. Map data from \cite{openstreetmap}}
    \label{fig:hwm_norain}
\end{figure}

High water marks (HWMs) are visible markings left by flood waters, such as debris lines, and are usually recorded shortly after a storm abates, in order to help better understand the flood and mitigate future risk. HWMs were not recorded by the USGS for this storm, but there were 104 collected in the Houston area by the Harris County Flood Control District \cite{hcfcd2024beryl}, ranging up to 2.5 meters above sea level. Of these, only two were located in ``wet'' parts of the mesh in the best track DG-SWEM model without rainfall forcing. By contrast, twenty-five HWMs were in wet elements in the compound flooding model. These twenty-five HWMs are and their relative errors are shown in Figures \ref{fig:hwm_rain} and \ref{fig:hwm_norain}.


\begin{figure}
    \centering
    \includegraphics[width=\linewidth]{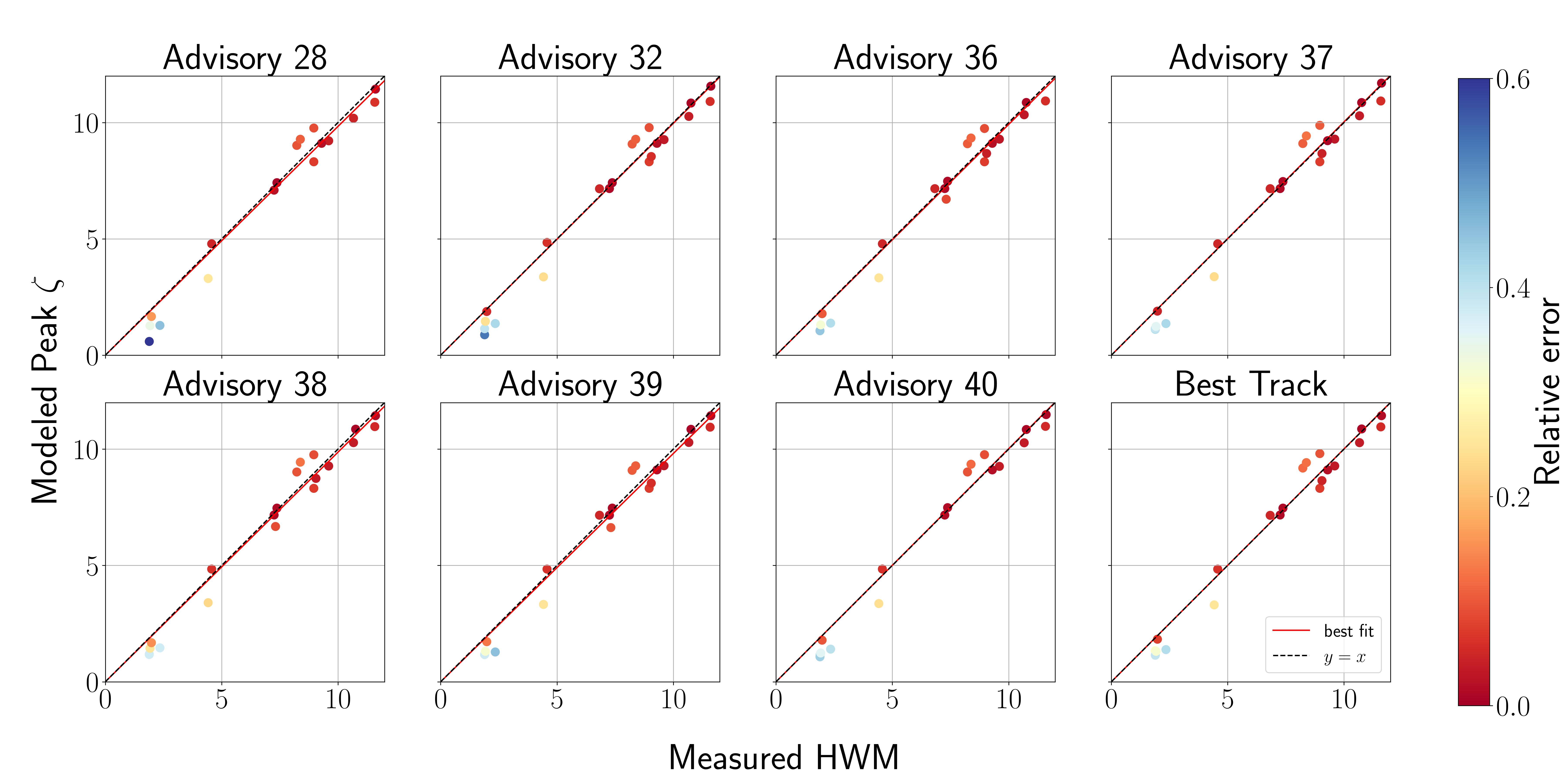}
    \caption{Peak surface elevations obtained with rainfall forcing, compared with observed HWM data}
    \label{fig:hwm_error_all}
\end{figure}

For this analysis and throughout this study, we calculate the RMSE and the best fit line through the origin $y = ax$. Producing slope $a$ close to $1$ indicates a highly accurate prediction model, on average. The high water marks for the best track hindcast, as well as each advisory are shown in \ref{fig:hwm_error_all}.

Some statistics from this data are summarized in Table \ref{tab:hwm_stats}. While the RMSE for these higher-elevation points is fairly high, for the best track hindcast, we find a best fit line through the origin with slope 1.0009. The relative error for the best track hindcast peaks at 0.41, but the majority of the wetted HWMs have relative errors under 10 percent, for each compound model run. The advisories' HWMs will be discussed further in Section \ref{sec:advisories}.

\begin{table}[ht]
\centering
\caption{HWM statistics for each simulation}
\begin{tabular}{l c c c l} 
\toprule
Simulation & Slope & $R^{2}$ & RMSE (m) & Count \\
\midrule
Advisory 28 & 0.9833 & 0.9573 & 0.6914 & 17 \\
Advisory 32 & 0.9965 & 0.9709 & 0.6069 & 24 \\
Advisory 36 & 0.9919 & 0.9704 & 0.5722 & 23 \\
Advisory 37 & 1.0029 & 0.9719 & 0.6004 & 25 \\
Advisory 38 & 0.9868 & 0.9739 & 0.5775 & 22 \\
Advisory 39 & 0.9811 & 0.9720 & 0.5655 & 22 \\
Advisory 40 & 1.0017 & 0.9739 & 0.6098 & 21 \\
Best Track  & 1.0009 & 0.9724 & 0.5955 & 25 \\
\bottomrule
\end{tabular} 
\label{tab:hwm_stats}
\end{table}

\subsection{Advisories} \label{sec:advisories}

In this section we review the results from each of seven modeled weather advisories, comparing the NOAA gauge data at selected weather stations with the simulated results. Each advisory was run with a timestep of 0.25 seconds, in order to maintain constistency between forecast models and to ensure stability when faced with rapid-onset strong winds and pressures. 

\begin{figure}
    \centering
    \includegraphics[width=0.5\linewidth]{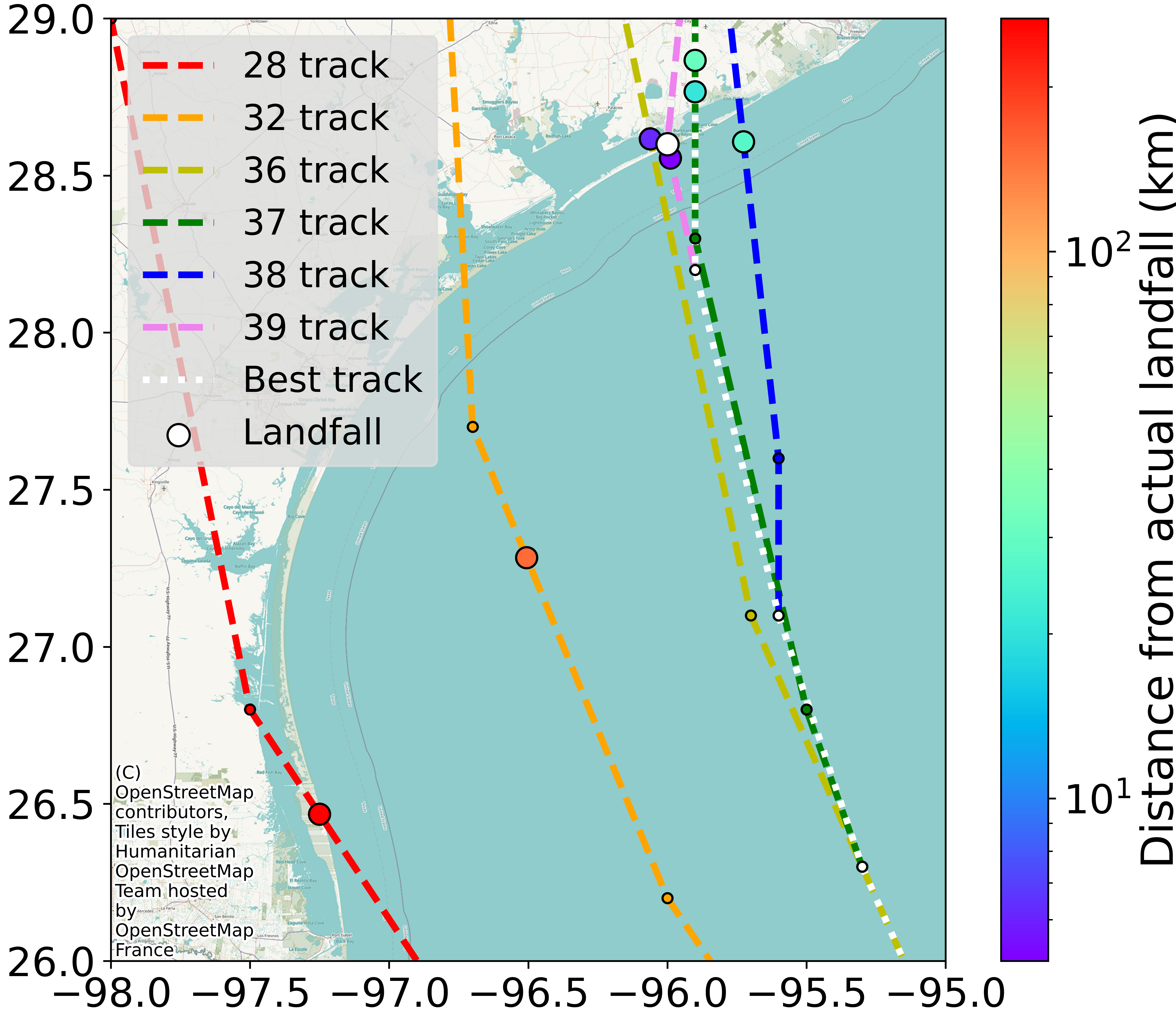}
    \caption{Interpolated storm centers (large circles) at the actual time of landfall for each advisory. Map from \cite{openstreetmap}}
    \label{fig:stormcenters}
\end{figure}

The forecasts vary significantly, being updated with new data from the NHC every few hours. As our model linearly interpolates the storm centers between each timestep, we can calculate the forecast locations of the storm centers for each advisory. In Figure \ref{fig:stormcenters}, we show the forecast storm centers at the actual time of landfall. We observe that the forecast vastly improved between Advisory 28 on July 5 and Advisory 36 on July 7.

Note that the interpolated paths of the best track and Advisory 37 are collinear as they pass over Matagorda Bay. The point closer to the landfall location actually comes from Advisory 37, while the interpolated best track data places the landfall slightly further away. The precise landfall location for each advisory is not necessarily crucial for each model to perform well; most crucially, here, we note the large discrepancies between the first two forecasts and the others.

\subsubsection{Advisory 28} \label{sec:adv28}

\begin{figure}
    \centering
    \includegraphics[width=\textwidth]{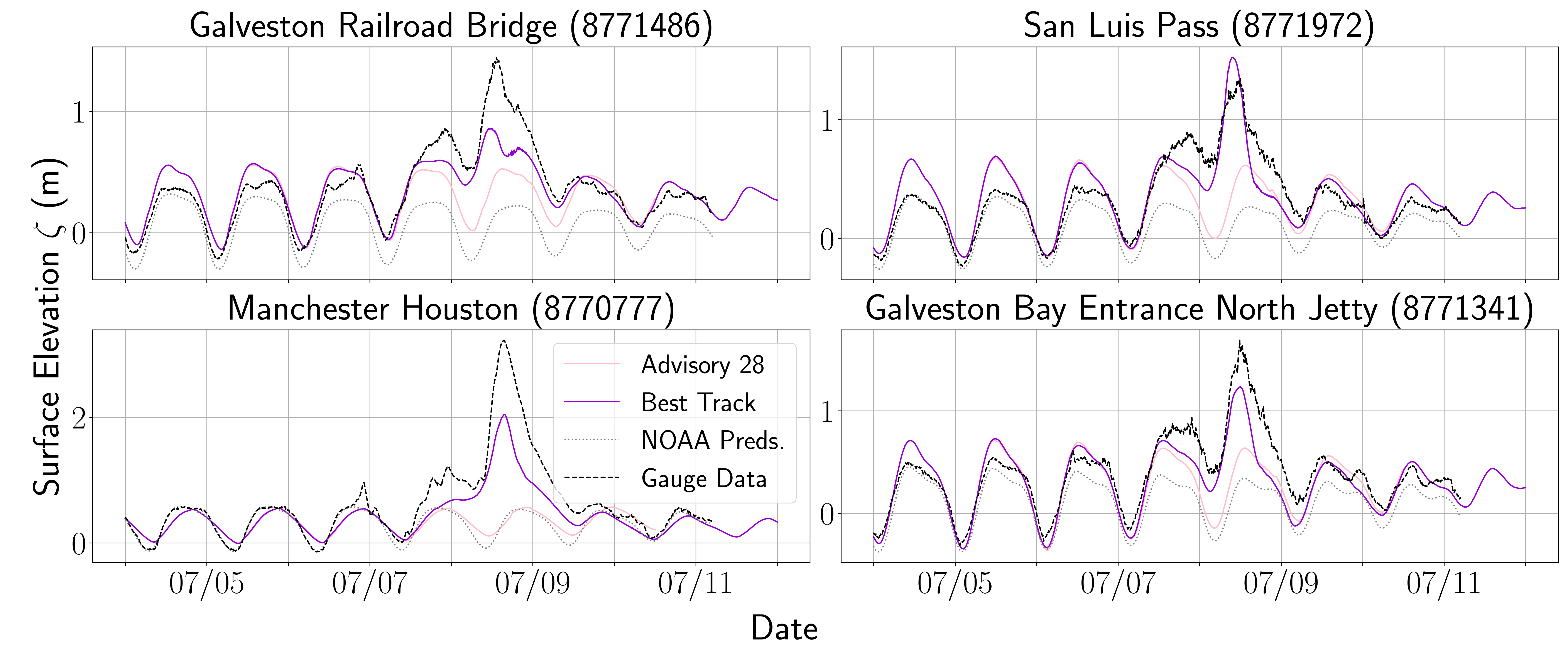}
    \caption{Forecast simulated from Advisory 28, compared with simulated best track hindcast and NOAA predictions and gauge data}
    \label{fig:advisory28}
\end{figure}

The first advisory we analyze was issued on July 5, three days before making landfall in the U.S., when the hurricane eye was over the Yucat\'an peninsula. Advisory 28 is a 5-day forecast, between July 5--10. Hurricane Beryl was forecast to proceed up the Gulf coast more slowly than it eventually did. As shown in Figure \ref{fig:stormcenters}, at the time of landfall, Advisory 28 had forecast the storm to be over 250 kilometers away, over South Padre Island. The model suffers from this inaccuracy, underpredicting the peak water surface elevations at each NOAA gauge, essentially following the same tidal pattern from before the storm.
The simulated results from Advisory 28 are shown in Figure \ref{fig:advisory28}, alongside the observed data from selected NOAA water elevation gauges and the simulated best track hindcast.

Without rainfall, this advisory resulted in none of the HWMs being ``wetted'' by the model; with rainfall, on the other hand, 17 HWMs were wet. This is the fewest of all the advisories. This can be explained by the predicted storm center being so far away from the actual landfall location, resulting in less rain and storm surge in the Houston area. Advisory 28 also resulted in the high water mark with the highest relative error, at 0.69, as shown in Figure \ref{fig:hwm_error_all}.

\subsubsection{Advisory 32}

\begin{figure}
    \centering
    \includegraphics[width=\textwidth]{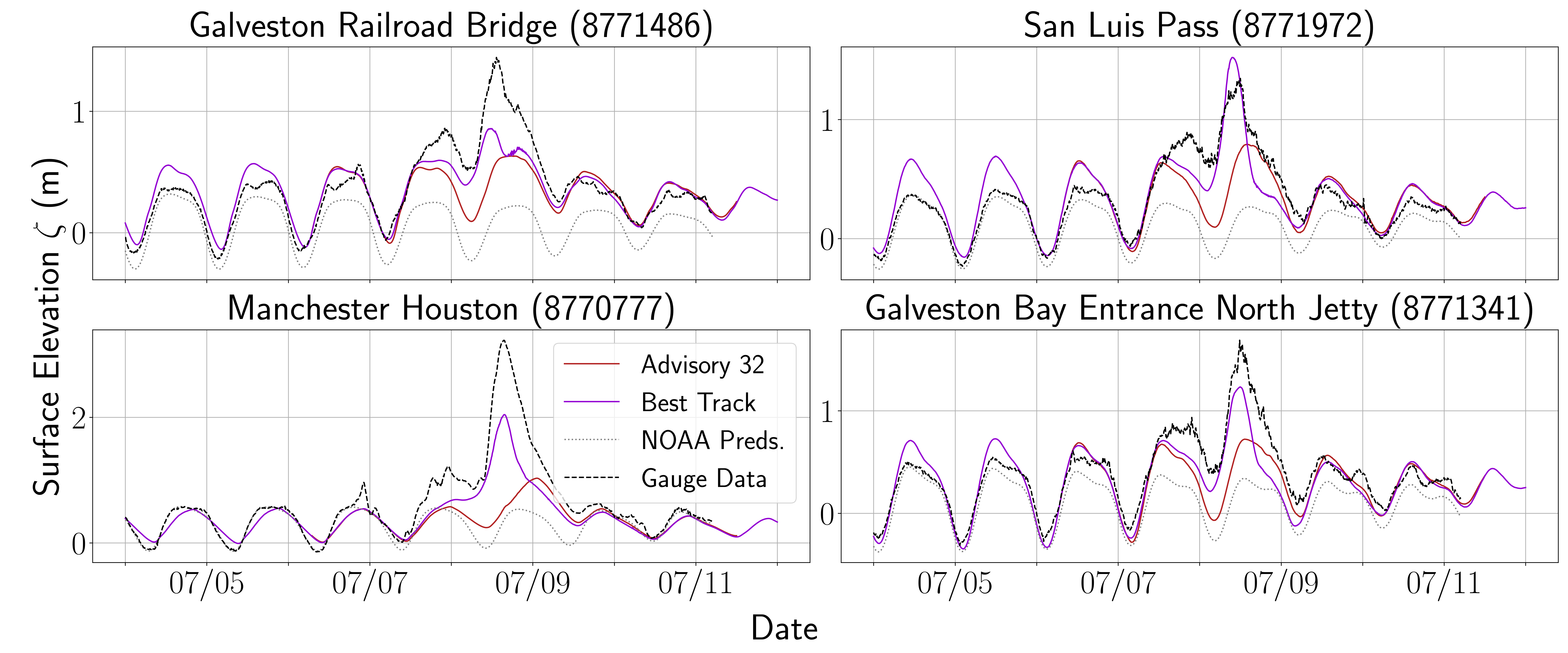}   
    \caption{Forecast simulated from Advisory 32, compared with simulated best track hindcast and NOAA predictions and gauge data}
    \label{fig:advisory32}
\end{figure}

Twenty four hours later, when Advisory 32 was issued, Beryl had moved northwest over the Gulf of Mexico, and the storm was downgraded to a tropical storm. Advisory 32 is another 5-day forecast, betwen July 6--11. This advisory had a forecast storm center southeast of Port Aransas at the time of landfall. This is much closer than Advisory 28, but still about 154 kilometers from Beryl's actual landfall location. The simulated results still suffer from this inaccuracy, but are improved slightly, showing a marked increase in peak water levels at the height of the storm.
The simulated results from Advisory 32 are shown in Figure \ref{fig:advisory32}.

Without rainfall, Advisory 32 and all subsequent advisories resulted in a single wet HWM, near Seabrook, Texas. With rainfall forcing added, this advisory contained the twenty-four HWMs, showing that even two days out from the storm making landfall, the surge-runoff interactions allow for modeling fairly far upstream. One HWM shows that the model penetrates as far inland as White Oak Bayou, a point which is not even present in the best track hindcast model.

We observe that the best fit line for Advisory 32 is significantly closer to $y = x$ than that of Advisory 28, and its RSME is slightly decreased, as shown in Table \ref{tab:hwm_stats}. This intuitively makes sense, as the the forecast storm center moved closer to its actual landfall location. 

\subsubsection{Advisory 36}

\begin{figure}
    \centering
    \includegraphics[width=\textwidth]{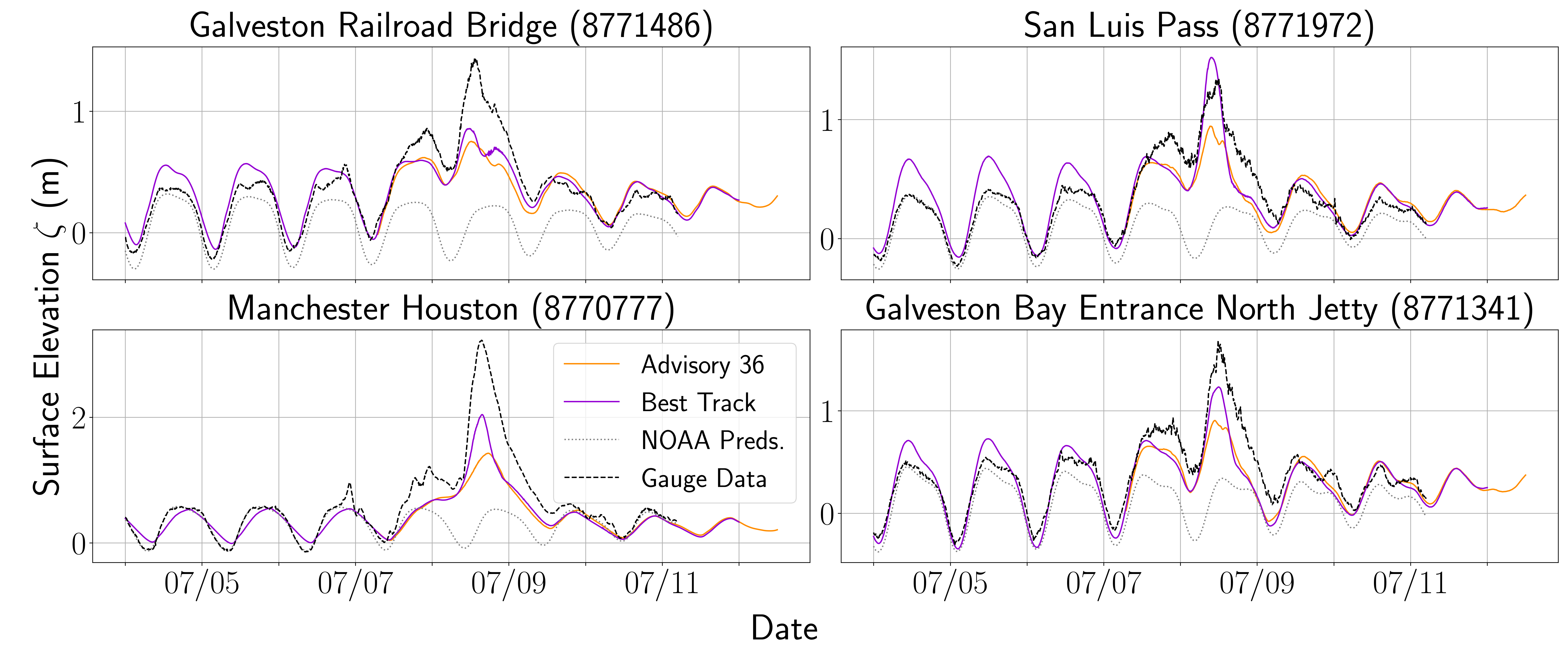}
    \caption{Forecast simulated from Advisory 36, compared with simulated best track hindcast and NOAA predictions and gauge data}
    \label{fig:advisory36}
\end{figure}

The next day, the storm center had moved north, east of Brownsville, when Advisory 36 was issued. Advisory 36 is another 5-day forecast, betwen July 7--12. This advisory's forecast storm center at the time of landfall was one of the closest to the actual location, at less than 10 kilometers away. This is a much more accurate forecast, being only a day away from landfall, and the simulated results reflect this. While the peaks still all underpredict the best track hindcast, they model the character of the gauge data very well.
The simulated results from Advisory 36 are shown in Figure \ref{fig:advisory36}.

In the compound model, Advisory 36 resulted in 23 wet HWMs, and a lower RMSE value than Advisory 32 and even the best track hindcast, as shown in Table \ref{tab:hwm_stats}. At this point, the forecast is fairly accurate, and the simulated runoff-surge interaction performs closer to what we observe in the measured HWM data.

\subsubsection{Advisory 37}

\begin{figure}
    \centering
    \includegraphics[width=\textwidth]{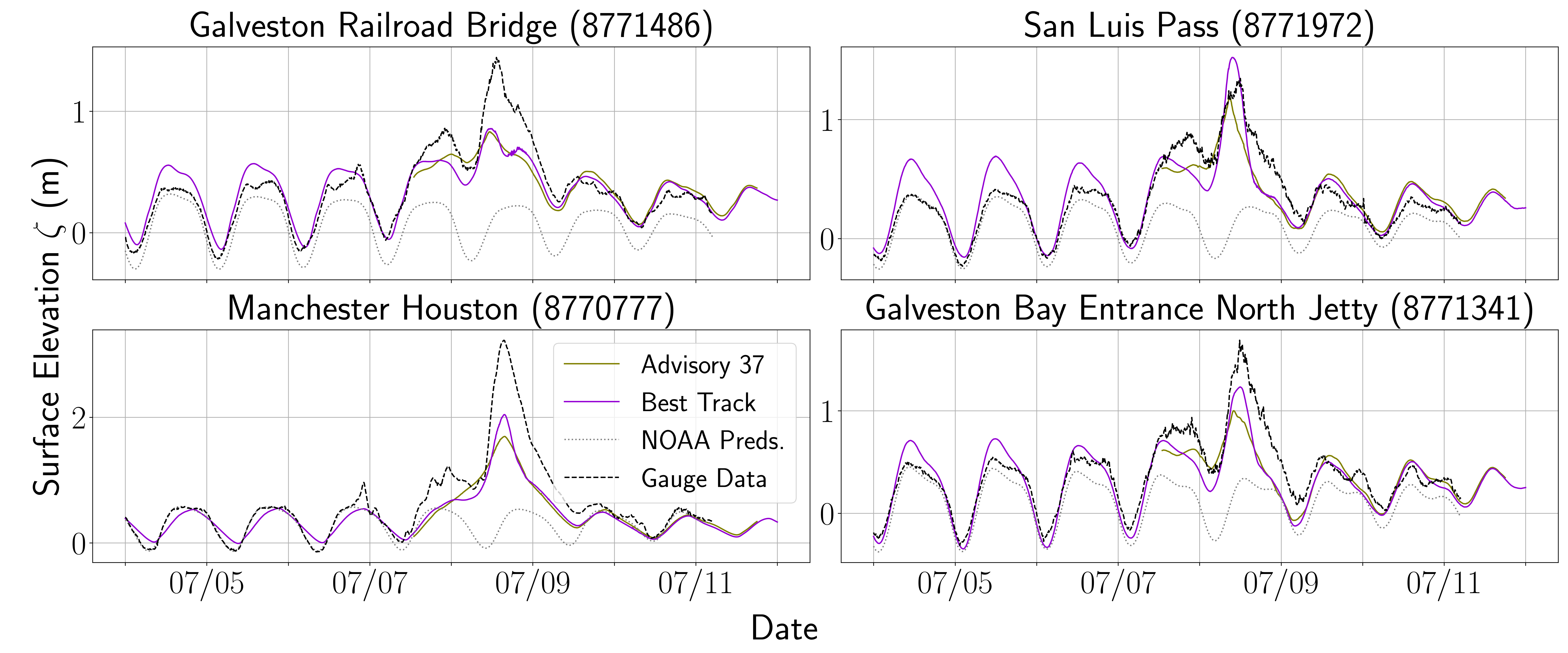}
    \caption{Forecast simulated from Advisory 37, compared with simulated best track hindcast and NOAA predictions and gauge data}
    \label{fig:advisory37}
\end{figure}

Now that the storm is within one day of landfall in Texas, we follow it at each 6-hour increment. Beryl continued on its northward trajectory in the Guld of Mexico. Advisory 37 is a 4--day forecast, between the evenings of July 7--11. Despite its forecast storm center at the time of landfall being slightly further away than that of the previous advisory, Advisory 37 significantly improves in its sampled results, closely approximating the best track hindcast for the selected gauges in Figure \ref{fig:advisory37}.

Advisory 37 resulted in 25 wet HWMs, the most of any advisory in this study. The RMSE increased slightly compared to the previous advisory, but the best fit line is yet closer to $y = x$, as shown in Table \ref{tab:hwm_stats}.

\subsubsection{Advisory 38}

\begin{figure}
    \centering
    \includegraphics[width=\textwidth]{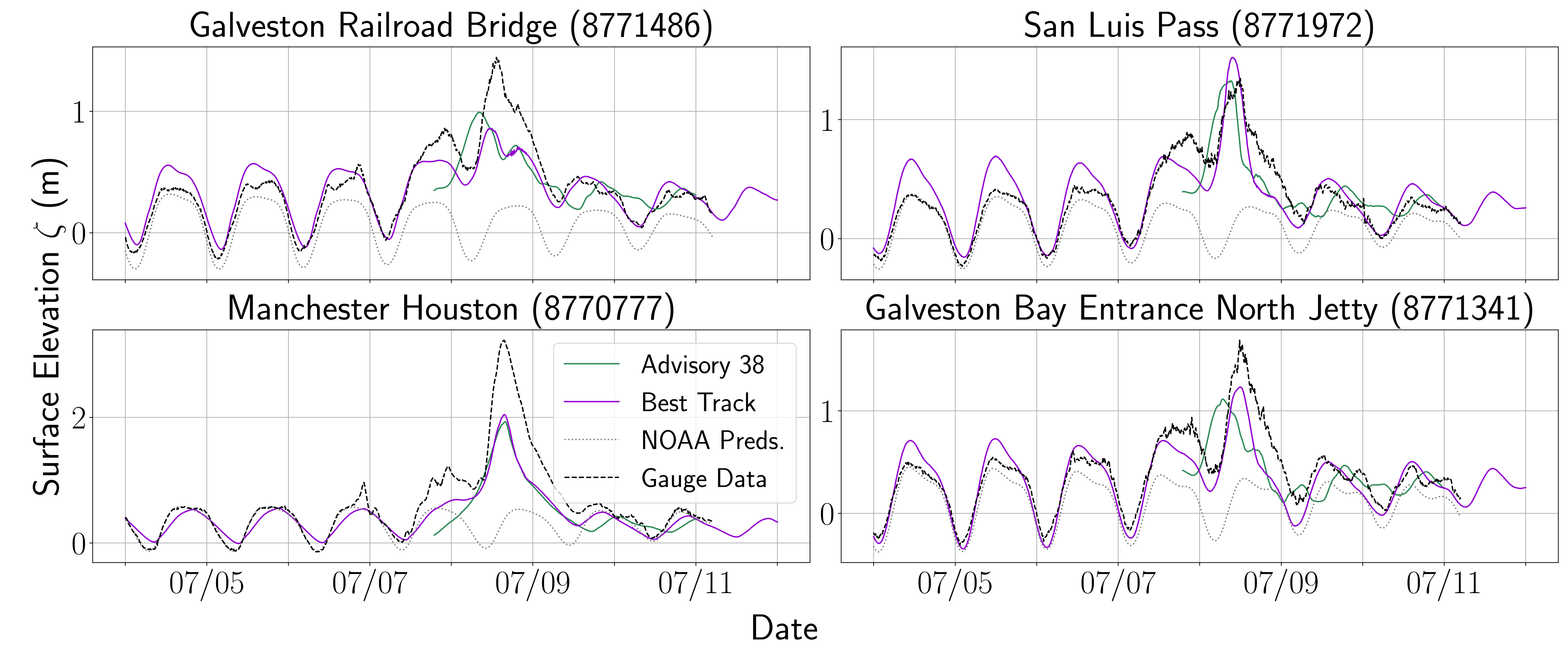}
    \caption{Forecast simulated from Advisory 38, compared with simulated best track hindcast and NOAA predictions and gauge data}
    \label{fig:advisory38}
\end{figure}

As Tropical Storm Beryl continued further north and Texans braced for impact, Advisory 38 was issued, now a 3-day forecast between July 8--11. Its forecast storm center at the time of landfall is still fairly close to Matagorda Bay, but slightly off to the east. The sampled points tell a strange story, predicting the peak surge a few hours before it hits, as shown in Figure \ref{fig:advisory38}. Interestingly, the peak surface elevation at Manchester Houston very closely approximates that of the best track hindcast, demonstrating appropriate inundation to inland regions despite the slightly offset modeled peaks outside of Galveston Bay. 

Advisory 38 resulted in 22 wet HWMs, with RMSE of 0.58 m. The coefficient of determination $R^{2}$ reaches a high, tying Advisory 39 as shown in Table \ref{tab:hwm_stats}.

\subsubsection{Advisory 39}

\begin{figure}
    \includegraphics[width=\textwidth]{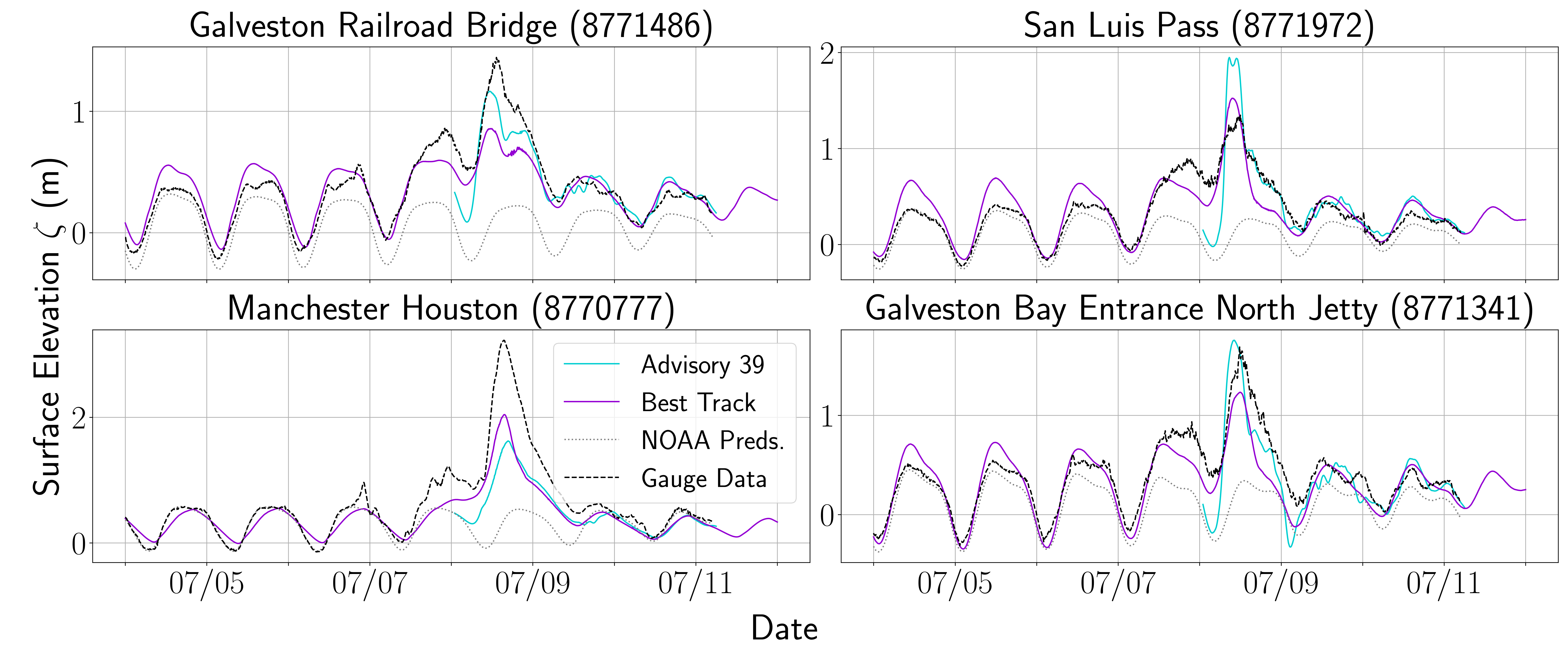}
    \caption{Forecast simulated from Advisory 39, compared with simulated best track hindcast and NOAA predictions and gauge data}
    \label{fig:advisory39}
\end{figure}

Now once again designated as a Category 1 Hurricane, Beryl continued northward. Advisory 39 was issued when Beryl was about 40 kilometers south of its eventual landfall location in Matagorda Bay. Advisory 39 is another 3-day forecast between July 8--11. Starting off with strong winds and an atmospheric pressure of 972 mbar, the model slightly overshoots the best track hindcast, very closely approximating the surface elevation gauges around Galveston Island, while still underpredicting at Manchester Houston, as shown in Figure \ref{fig:advisory39}.

This advisory resulted in 22 wet HWMs, with the lowest RMSE of all the models, at 0.57 m, as shown in Table \ref{tab:hwm_stats}.

\subsubsection{Advisory 40}  \label{sec:adv40}

\begin{figure}
    \centering
    \includegraphics[width=\textwidth]{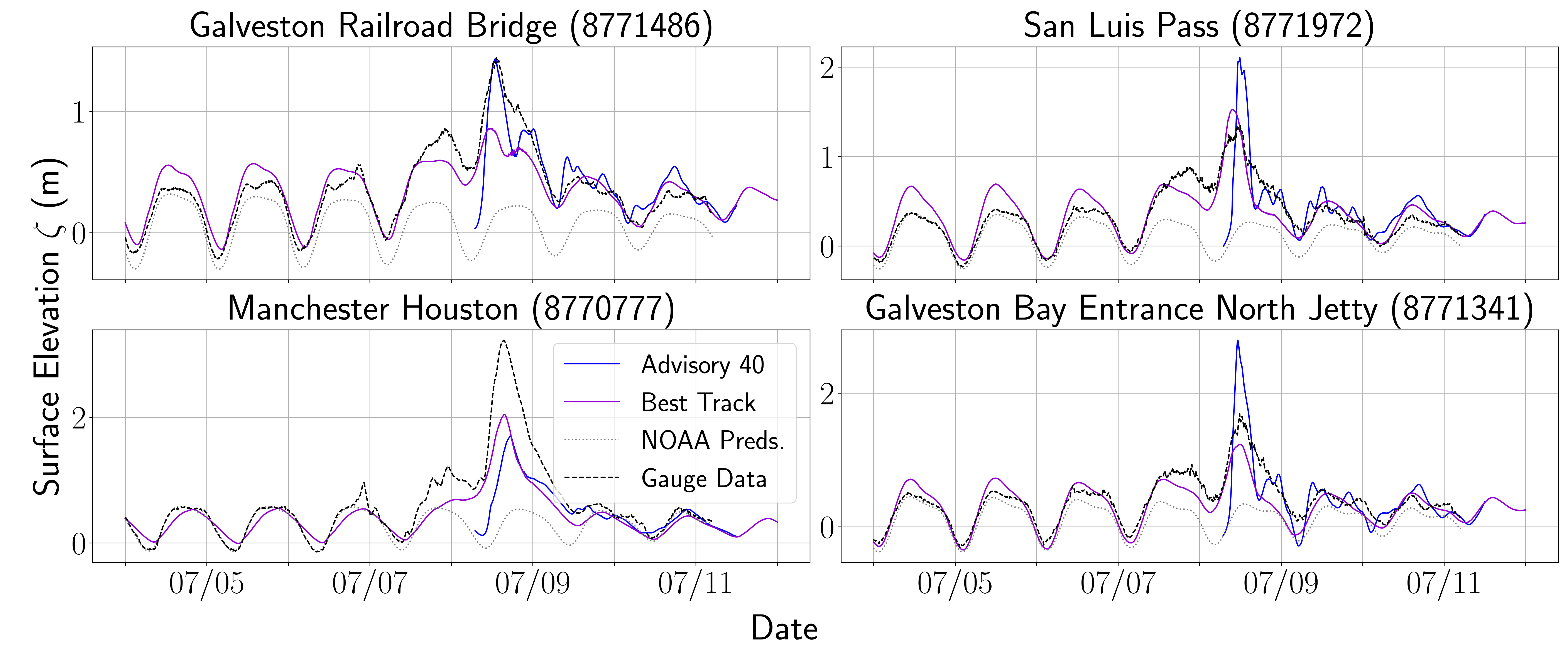}
    \caption{Forecast simulated from Advisory 40, compared with simulated best track hindcast and NOAA predictions and gauge data}
    \label{fig:advisory40}
\end{figure}

For Advisory 40, Beryl was again downgraded to a tropical storm as it continued north through Houston. Advisory 40 is another shorter 3-day forecast between July 8--11. 
The simulated results from Advisory 40 are shown in Figure \ref{fig:advisory40}. This advisory, issued after landfall in Texas, begins with strong winds right away, and the water surface elevation in many areas quickly rises to a high peak, overpredicting some of the gauge data. These peaks also overshoot those of the best track hindcast in the areas more open to the ocean, but undershoot further upstream at the Manchester Houston gauge.

Like Advisories 32 and 36, without rainfall added, this advisory resuluted in one a single wet HWM, near Seabrook, Texas. With rainfall, it resulted in 21 wet HWMs, but the RMSE increased compared to Advisory 36, as shown in Table \ref{tab:hwm_stats}. This increase is due, in part, to the quick jump at the beginning of the model, as will be discussed in Section \ref{sec:besttrack}. Nonetheless, Advisory 40 resulted in the highest $R^{2}$ value and the best fit line closest to $y=x$, which attests to the high accuracy of the more recently updated forecast.

\subsection{Best Track Analysis} \label{sec:besttrack}

We also sample many different points along the Texas coast throughout each model, and compare the peaks of these sampling points from each advisory, with and without rain, to the best track hindcast model's data.
Note that while the best track hindcast was also run both with and without rainfall forcing, the one used in this comparison includes the rainfall forcing term; this was chosen because of the conclusions from Section \ref{sec:validation}, which show slightly higher overall accuracy and much more upstream inundation for the compound flooding model. 

\begin{figure}
    \centering
    \includegraphics[width=0.9\textwidth]{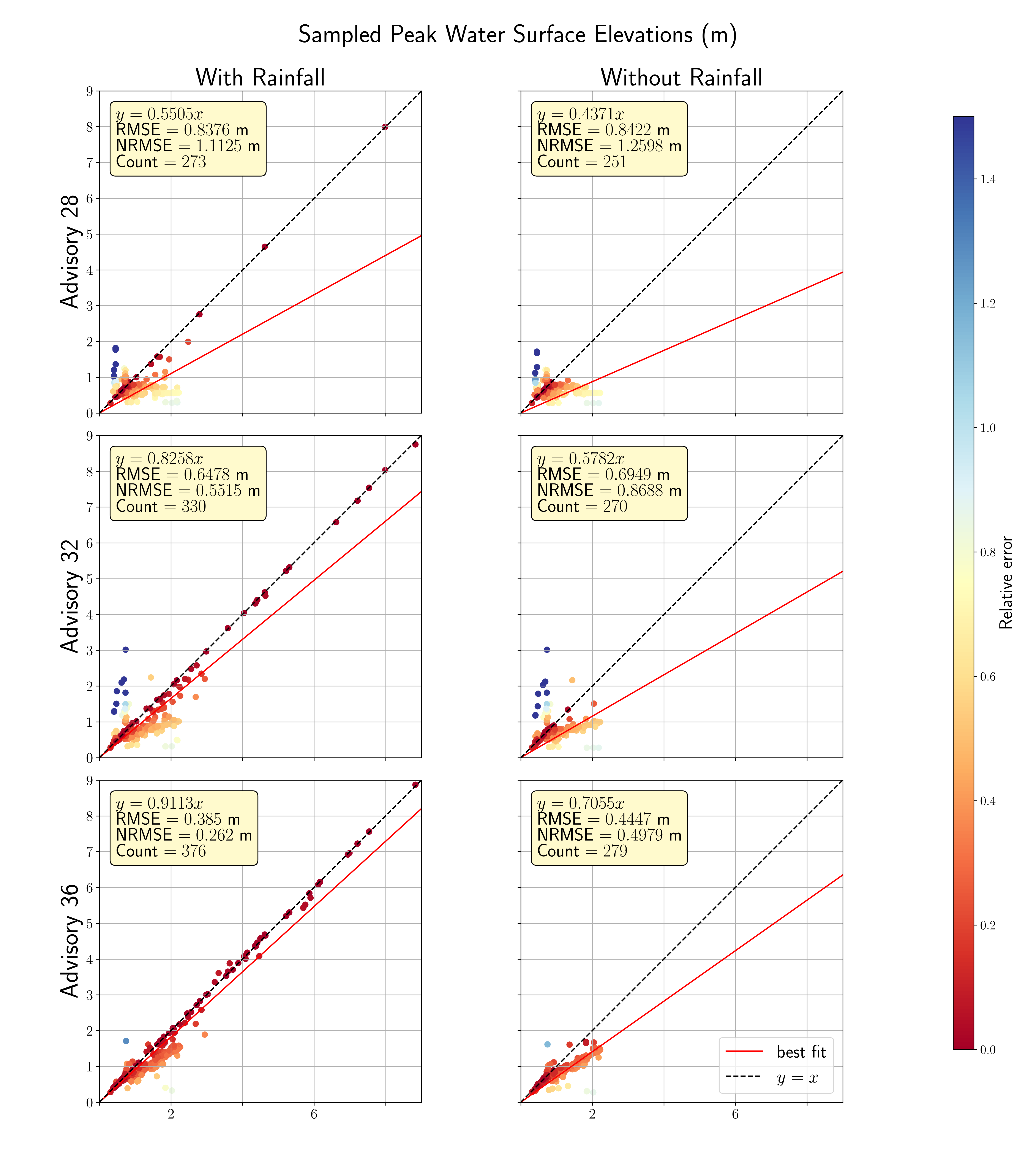}
    \caption{Relative error between the results of each advisory and the best track hindcast, for Advisories 23, 32, and 36}
    \label{fig:besttrack_err1}
\end{figure}

\begin{figure}
    \centering
    \includegraphics[width=0.85\textwidth]{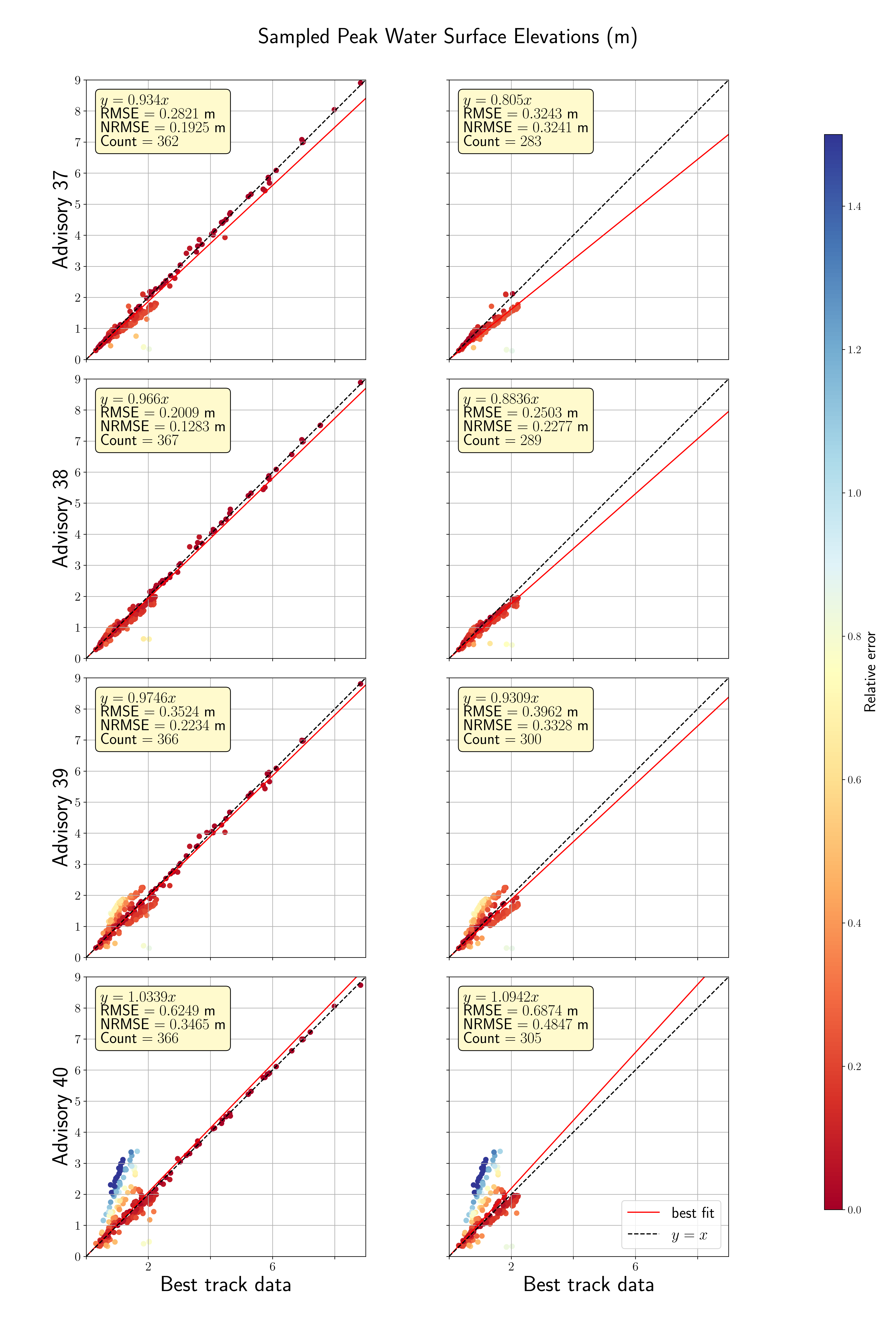}
    \caption{Relative error between the results of each advisory and the best track hindcast, for Advisories 37--40}
    \label{fig:besttrack_err2}
\end{figure}

The sampled peak elevations and relative errors are shown in Figures \ref{fig:besttrack_err1} and \ref{fig:besttrack_err2}. A total of 471 points were sampled for each model, but only the wet nodes for each model are shown in the Figures. 

When compared with the best track hindcast, the forecasts again perform successively better with each advisory.
For the compound flooding model, the best fit line improves dramatically, from a slope of 0.5505 for Advisory 28 to a slope of 1.0339 for Advisory 40. Advisories 28 and 32 contain many points that underpredict the best track hindcast, but several outliers that overpredict it by a large margin, as shown in Figure \ref{fig:besttrack_err1}. Advisories 36--38 do not contain these outliers, instead slightly underpredicting in nearly all the sampled points. Advisory 39 models many points slightly higher, and many others slightly lower, than the best track hindcast. Advisory 40, on the other hand, has some peaks far higher than the best track, as was demonstrated with some elevation gauges in Figure \ref{fig:advisory40}. These skew the best fit line up, and their high relative errors cause the RMSE to increase, though it had been decreasing for the first three advisories.

Looking at the models without rainfall added, we observe similar patterns. The RMSE decreases for the first three advisories, and then jumps back up for Advisory 40 due to some heavy overpredictions. The best fit also approaches $y=x$, albeit significantly more slowly.

However, when we compare this simpler model with the compound flooding one, we observe major differences in the distributions of peak elevations. The compound model's sampled datapoints are distributed over a wide range of peak elevations, because of the traditional model's lack of wet data points in regions of higher elevation. The points unique to the compound model are all very close to the best track hindcast, with relative errors less than 0.1. This fact alone does not prove that the additional rainfall substantially increases flooding in these upstream areas; that was already demonstrated in Figure \ref{fig:diff_map}. Rather, it shows that the compound model captures more upstream channels because of its ability to wet nodes via rainfall, thus expanding the effective modeling domain.

In order to take into account for the large variance between the traditional and compound models, a normalized root mean square error (NRMSE) is computed as in Equation \ref{eq:nrmse}. 

\begin{align}
    \text{NRMSE} &= \frac{\text{RMSE}}{\bar{\zeta}^{\text{max}}_{a}} \label{eq:nrmse}
\end{align}
where $\bar{\zeta}^{\text{max}}_{a}$ is the mean of the sampled peak surface elevations for advisory $a$.

While the RMSE remains somewhat close between the two models, the NRMSE is remarkably lower for the compound flooding model. This improvement that increases as more inland elements are wetted by the rainfall, which drives up the mean peak elevations.

\section{Concluding Remarks} \label{sec:conclusions}

In this work, we analyze the forecasting potential of a new compound flooding model, which adds rainfall runoff to the long-established storm surge model, for an example case in Hurricane Beryl. It is shown that for this case, the rainfall forcing term from R-CLIPER greatly improves the model's ability to inundate low-lying inland areas, even with moderate storm surge. The best track hindcast is shown to increase the water column height by upwards of 50 centimeters in many low-lying wetland regions, especially near and along rivers.

Seven forecasting advisories were simulated both with and without rainfall, and compared to water elevation gauge data, high water marks in Harris County, Texas, and other simulated results from the observed best track hindcast. These advisories vary greatly in their predictions, with Advisory 28's predicted storm center lying over 200 kilometers away at the time of landfall, and each subsequent advisory approaching the best track slightly more.

Using NOAA water elevation gauge data, it is shown that the earlier selected advisories' models severely underpredict the peak elevations recorded by the weather stations. Over time, as the forecasts become more accurate, the modeled peaks more closely approximate the gauge data, with Advisories 39 and 40 even overpredicting at some stations along Galveston Island and Bolivar Peninsula. 

The few available HWMs, almost all missed by the surge-only model, show that each subsequent advisory more closely approximates the measured data at these upstream points. Even so, the RMSE at these points remains fairly high, at over 0.5 meters. Modeling a storm better recorded would provide a more statistically significant result here, since generally a flood of this magnitude would have better data available (c.f. Hurricane Harvey (2017) which has 2620 available HWMs \cite{floodeventviewer}).

When the maximum elevation of 471 points, sampled throughout each advisory's simulation, are compared to those of best track hindcast, the extent of the rainfall runoff term's improved inundation is more explicitly shown. The rainfall-free model hardly penetrates above two meters, while the compound model contains dozens of additional datapoints, at elevations up to nine meters. Though in many deeper areas the rainfall had minimal contribution, its addition allows the model to permeate much deeper inland to areas that became inundated by Beryl's flooding. This increased modeling domain is of great potential use to flood plain analysis and predictions.

Applying the relatively simple parametric rainfall term to this forecasting scenario is shown to result in vastly more descriptive results at each advisory, despite the vast difference in storm center forecasts. Many more upstream points were able to be modeled, and the NRMSE in particular shows the great improvements gained using compound flooding. 

There are several river stage gauges in the area impacted by Hurricane Beryl, which exhibit much greater surface elevation increases during Beryl. For example, the Tres Palacios River near Midfield, Texas (in Matagorda County, some 25 kilometers inland from Beryl's landfall) increased in stage by more than 7.5 meters \cite{usgstrespalacios}. Unfortunately, these are near or just beyond the extents of the current mesh domain, limiting their applicability in this study. Expanding the domain further inland to allow incorporation of these river gauges would improve future analysis capability. 

Furthermore, in order to even more effectively model inland flooding, the model could also be forced by river discharge, applied further upstream than the added rainfall, as shown in \cite{loveland2021developing, wichitrnithed2024discontinuous}. This would likely further improve the model along rivers and in low-lying regions. 

\section*{Acknowledgments}

This work has been supported by the U.S. National Aeronautics and Space Administration (NASA) under Grant Award No. 80NSSC24K1505. The authors also would like to gratefully acknowledge the use of the ``ADCIRC,'' ``DMS23001,'' and ``DMS21031'' allocations on the Frontera supercomputer at the Texas Advanced Computing Center at the University of Texas at Austin.

\appendix

\newpage

\section{Simulated Advisories at Gauges}\label{appendix:advisories}

These plots show simulated results from each advisory compared with best track simulation and NOAA gauge data and predictions, a few days before and after Beryl's landfall in the United States.

\begin{figure}[ht]
    \centering
    \includegraphics[width=0.7\textwidth]{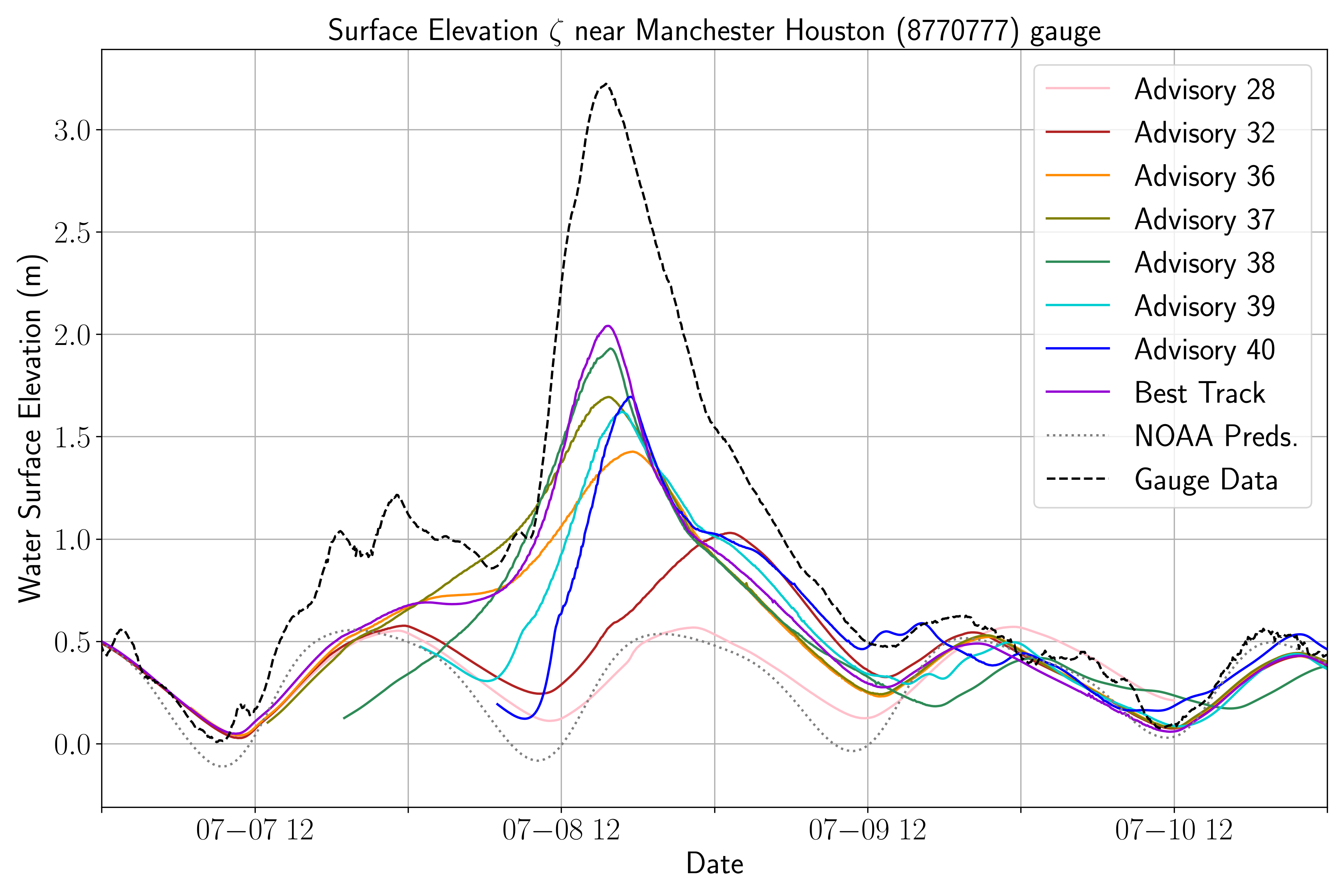}
    \caption{Manchester Houston (NOAA gauge 8770777)}
    \label{fig:manchester}
\end{figure}

\begin{figure}[ht]
    \centering
    \includegraphics[width=0.7\textwidth]{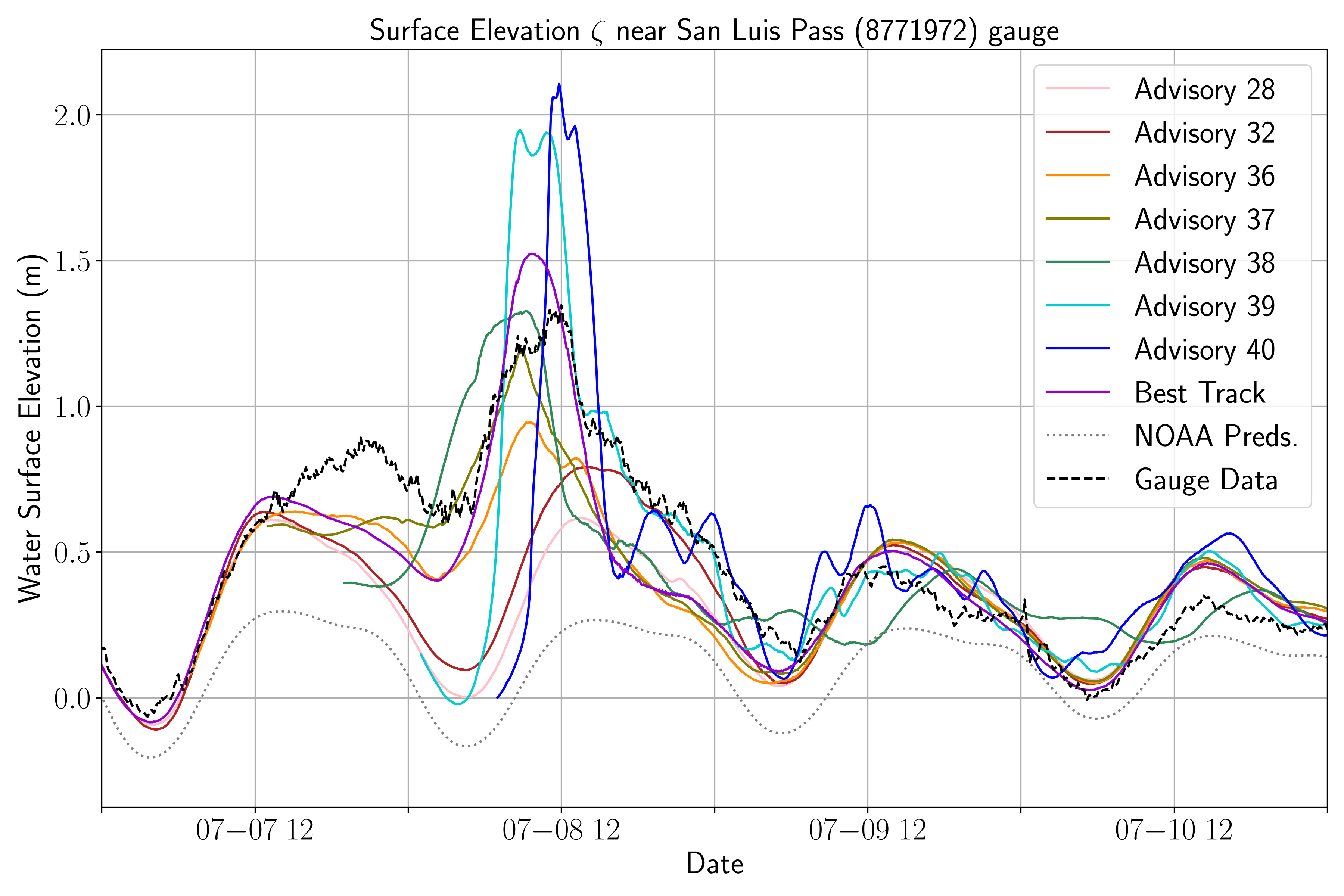}
    \caption{San Luis Pass (NOAA gauge 8771972)}
    \label{fig:sanluis}
\end{figure}

\pagebreak

\begin{figure}[ht]
    \centering
    \includegraphics[width=0.7\textwidth]{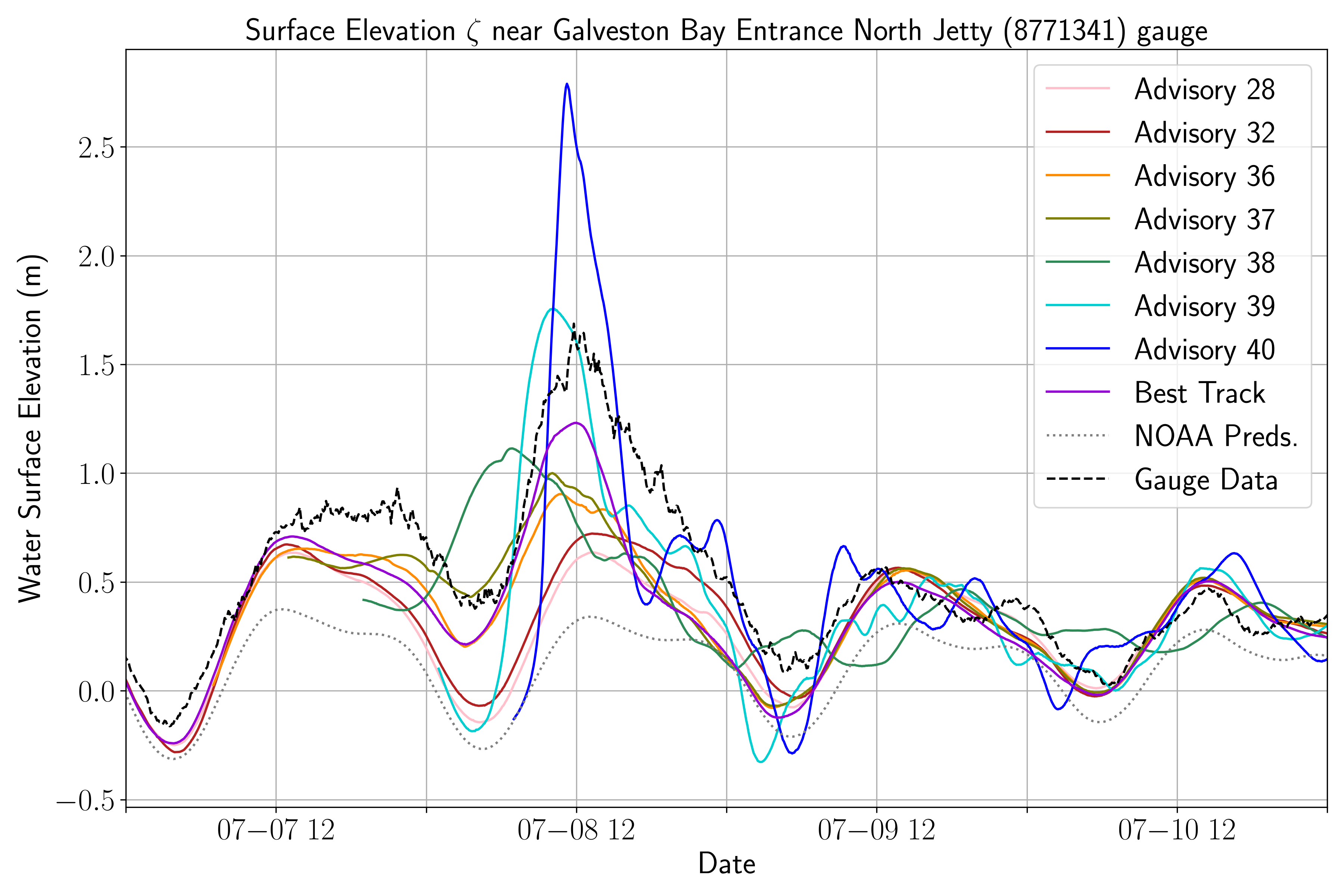}
    \caption{Galveston Bay Entrance North Jetty (NOAA gauge 8771341)}
    \label{fig:galvestonbay}
\end{figure}

\begin{figure}[ht]
    \centering
    \includegraphics[width=0.7\textwidth]{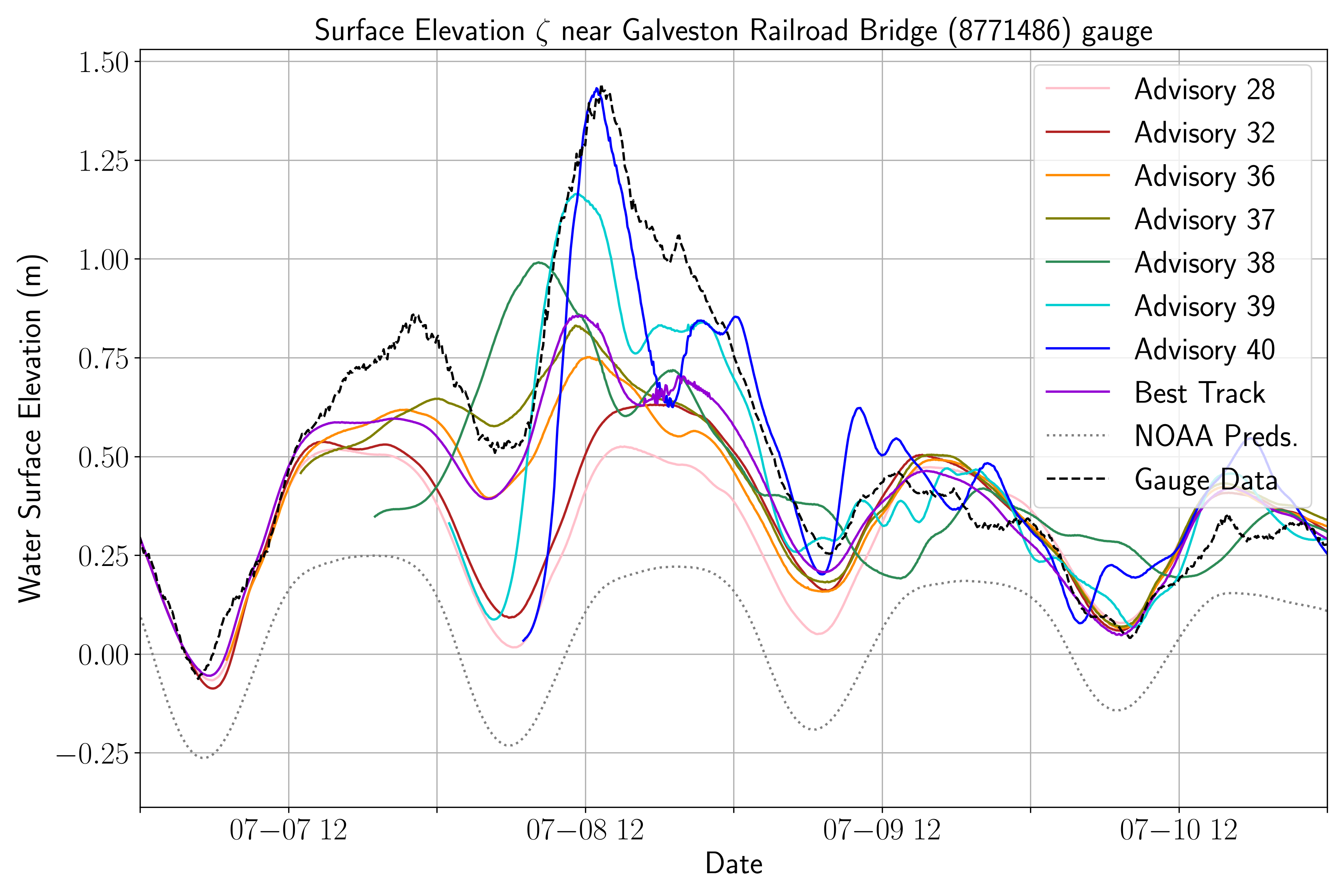}
    \caption{Galveston Railroad Bridge (NOAA gauge 8771486)}
    \label{fig:galvestonrr}
\end{figure}

\pagebreak 

\begin{figure}[ht]
    \centering
    \includegraphics[width=0.7\textwidth]{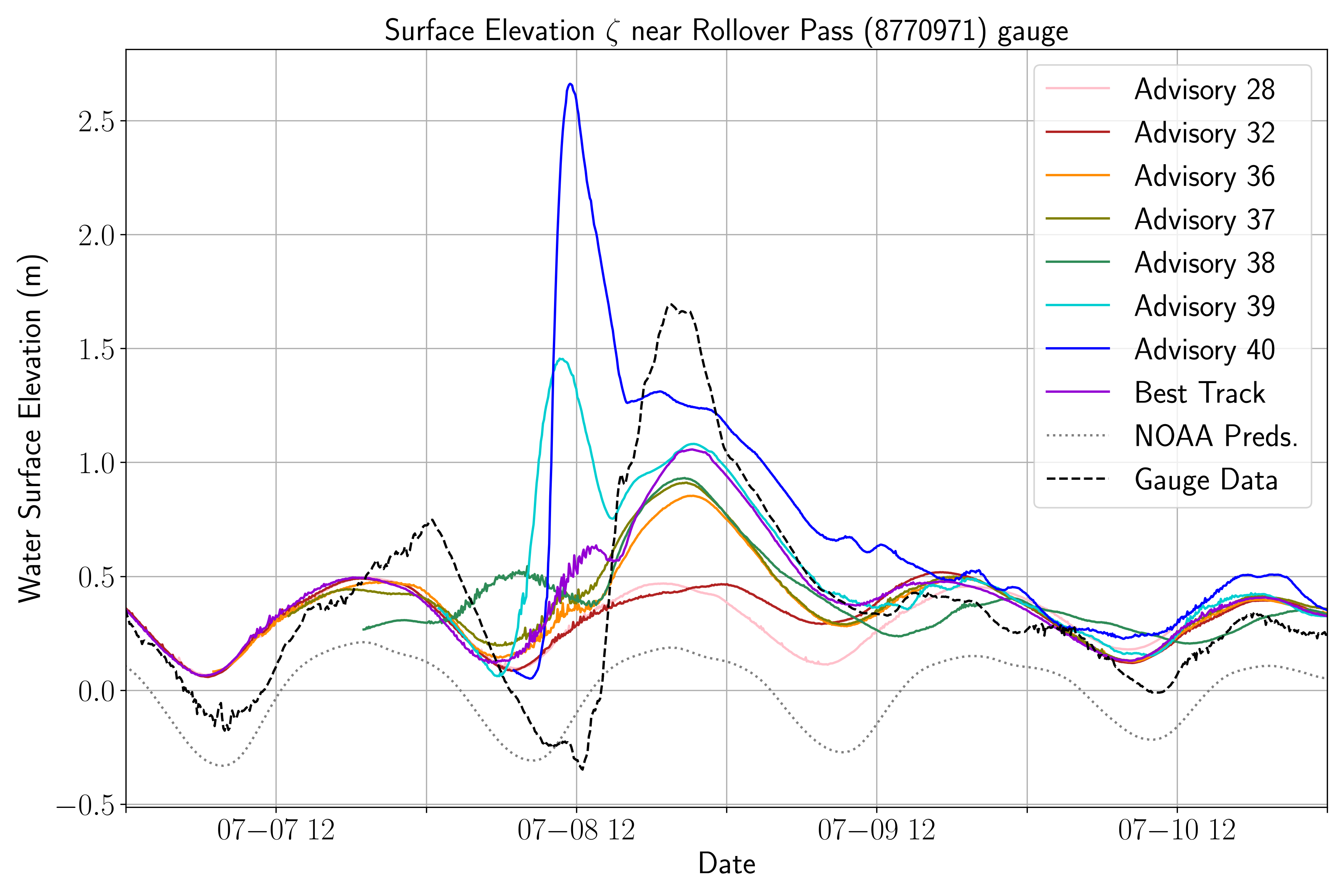}
    \caption{Rollover Pass (NOAA gauge 8770971)}
    \label{fig:rollover}
\end{figure}

\begin{figure}[ht]
    \centering
    \includegraphics[width=0.7\textwidth]{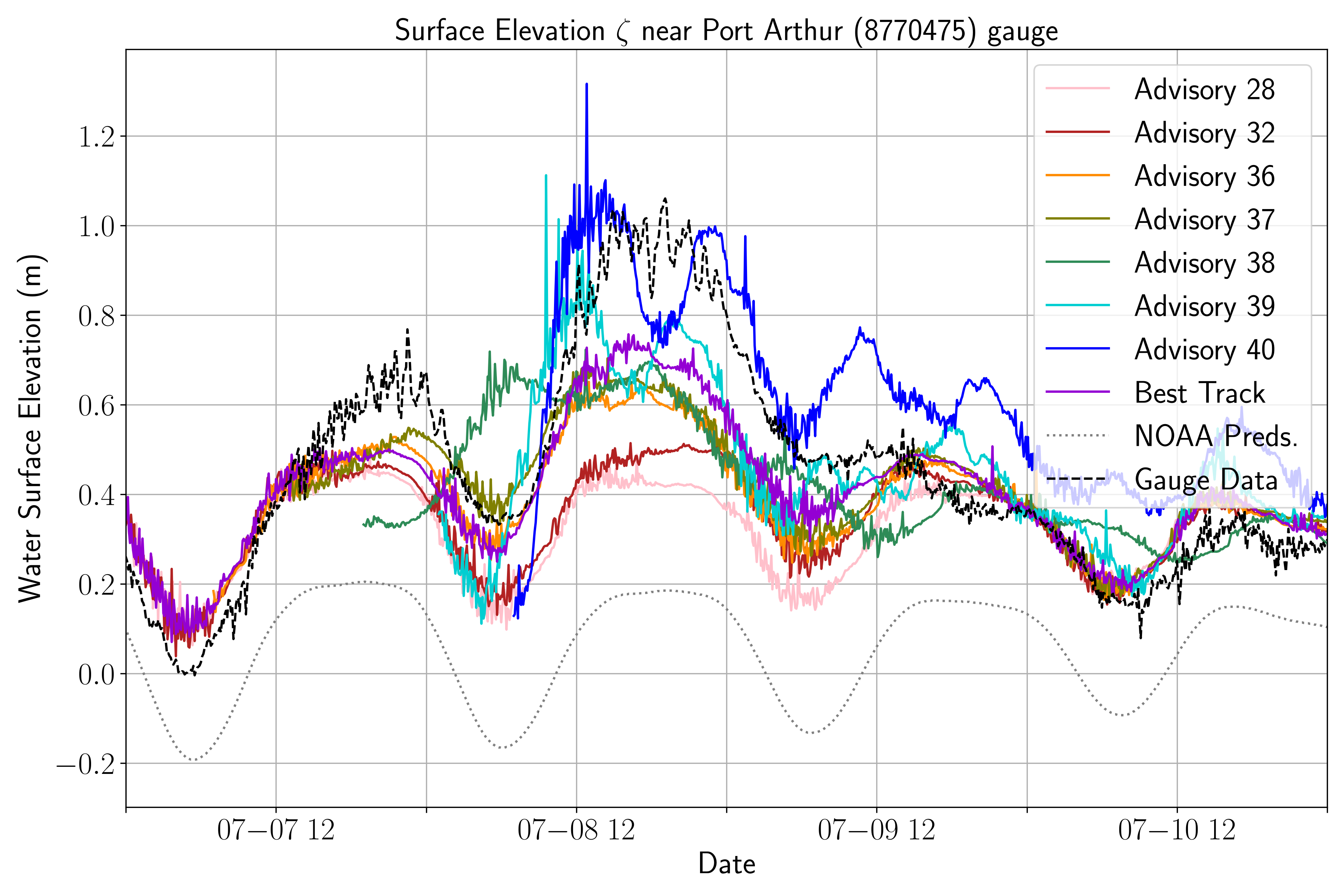}
    \caption{Port Arthur (NOAA gauge 8770475)}
    \label{fig:portarthur}
\end{figure}

\pagebreak

\bibliographystyle{elsarticle-num}
\bibliography{references}

@techreport{OConnor2001,
  title={Service Assessment: {Tropical Storm Allison}, Heavy Rains and Floods, {Texas and Louisiana, June 2001}},
  author={O'Connor, William and others},
  institution={U.S. Department of Commerce, NOAA/National Weather Service},
  year={2001},
  address={Silver Spring, MD},
}

@techreport{harvey2017,
  author      = "Blake, Eric and Zelinsky, David",
  title       = "Tropical Cyclone Report, {Hurricane Harvey (AL092017)}",
  institution = "National Hurricane Center",
  year        = "2008"
}

@techreport{ike2008,
  author      = "Berg, Robbie",
  title       = "Tropical Cyclone Report, {Hurricane Ike (AL092008)}",
  institution = "National Hurricane Center",
  year        = "2008"
}

@ARTICLE{Brackins2020-xp,
  title     = "Evaluation of parametric precipitation models in reproducing
               tropical cyclone rainfall patterns",
  author    = "Brackins, John T and Kalyanapu, Alfred J",
  journal   = "J. Hydrol. (Amst.)",
  publisher = "Elsevier BV",
  volume    =  580,
  number    =  124255,
  pages     =  124255,
  month     =  jan,
  year      =  2020,
  url       = "http://dx.doi.org/10.1016/j.jhydrol.2019.124255",
  language  = "en"
}

@TECHREPORT{Contreras2023-qi,
  title       = "A {Channel-to-Basin} Scale {ADCIRC} Based Hydrodynamic
                 Unstructured Mesh Model for the {US} {East and Gulf of Mexico}
                 Coasts",
  author      = "Contreras, Maria Teresa and Woods, Brendan and Blakely,
                 Coleman and Wirasaet, Damrongsak and Westerink, Joannes and
                 Cobell, Zach and Pringle, William and Moghimi, Saeed and
                 Vinogradov, Sergey and Myers, Edward and Seroka, Greg and
                 Lalime, Michael and Funakoshi, Yuji and Van der Westhuysen,
                 Andre and Abdolali, Ali and Valseth, Eirik and Dawson, Clint",
  number      = "NOS CS 54",
  institution = "National Oceanic and Atmospheric Administration",
  month       =  jan,
  year        =  2023,
  url         = "https://repository.library.noaa.gov/view/noaa/48079/noaa_48079_DS1.pdf"
}

@article{wahl2015increasing,
  title={Increasing risk of compound flooding from storm surge and rainfall for major {US} cities},
  author={Wahl, Thomas and Jain, Shaleen and Bender, Jens and Meyers, Steven D and Luther, Mark E},
  journal={Nature Climate Change},
  volume={5},
  number={12},
  pages={1093--1097},
  year={2015},
  publisher={Nature Publishing Group}
}

@ARTICLE{Kubatko2014-nc,
  title    = "Optimal {Strong-Stability-Preserving} {Runge--Kutta} Time
              Discretizations for Discontinuous {Galerkin} Methods",
  author   = "Kubatko, Ethan J and Yeager, Benjamin A and Ketcheson, David I",
  journal  = "J. Sci. Comput.",
  volume   =  60,
  number   =  2,
  pages    = "313--344",
  month    =  aug,
  year     =  2014
}

@ARTICLE{Shu1988-eq,
   title    = "Efficient implementation of essentially non-oscillatory
               shock-capturing schemes",
   author   = "Shu, Chi-Wang and Osher, Stanley",
   journal  = "J. Comput. Phys.",
   volume   =  77,
   number   =  2,
   pages    = "439--471",
   month    =  aug,
   year     =  1988
 }

@article{loveland2021developing,
  title={Developing a Modeling Framework to Simulate Compound Flooding: When Storm Surge Interacts With Riverine Flow},
  author={Loveland, Mark and Kiaghadi, Amin and Dawson, Clint N and Rifai, Hanadi S and Misra, Shubhra and Mosser, Helena and Parola, Alessandro},
  journal={Frontiers in Climate},
  volume={2},
  pages={609610},
  year={2021},
  publisher={Frontiers Media SA}
}

@article{santiago2019comprehensive,
  title={A comprehensive review of compound inundation models in low-gradient coastal watersheds},
  author={Santiago-Collazo, F{\'e}lix L and Bilskie, Matthew V and Hagen, Scott C},
  journal={Environmental Modelling \& Software},
  volume={119},
  pages={166--181},
  year={2019},
  publisher={Elsevier}
}

@article{orton2020flood,
  title={Flood hazard assessment from storm tides, rain and sea level rise for a tidal river estuary},
  author={Orton, PM and Conticello, FR and Cioffi, F and Hall, TM and Georgas, N and Lall, U and Blumberg, AF and MacManus, K},
  journal={Natural hazards},
  volume={102},
  number={2},
  pages={729--757},
  year={2020},
  publisher={Springer}
}

@article{kumbier2018investigating,
  title={Investigating compound flooding in an estuary using hydrodynamic modelling: a case study from the {Shoalhaven River}, {Australia}},
  author={Kumbier, Kristian and Carvalho, Rafael C and Vafeidis, Athanasios T and Woodroffe, Colin D},
  journal={Natural Hazards and Earth System Sciences},
  volume={18},
  number={2},
  pages={463--477},
  year={2018},
  publisher={Copernicus GmbH}
}

@article{kubatko2006hp,
  title={hp discontinuous {Galerkin} methods for advection dominated problems in shallow water flow},
  author={Kubatko, Ethan J and Westerink, Joannes J and Dawson, Clint},
  journal={Computer Methods in Applied Mechanics and Engineering},
  volume={196},
  number={1-3},
  pages={437--451},
  year={2006},
  publisher={Elsevier}
}

@article{dawson2011discontinuous,
  title={Discontinuous {Galerkin} methods for modeling hurricane storm surge},
  author={Dawson, Clint and Kubatko, Ethan J and Westerink, Joannes J and Trahan, Corey and Mirabito, Christopher and Michoski, Craig and Panda, Nishant},
  journal={Advances in Water Resources},
  volume={34},
  number={9},
  pages={1165--1176},
  year={2011},
  publisher={Elsevier}
}

@article{bunya2009wetting,
  title={A wetting and drying treatment for the {Runge--Kutta discontinuous {Galerkin}} solution to the shallow water equations},
  author={Bunya, Shintaro and Kubatko, Ethan J and Westerink, Joannes J and Dawson, Clint},
  journal={Computer Methods in Applied Mechanics and Engineering},
  volume={198},
  number={17-20},
  pages={1548--1562},
  year={2009},
  publisher={Elsevier}
}

@book{tan1992shallow,
  title={Shallow water hydrodynamics: Mathematical theory and numerical solution for a two-dimensional system of shallow-water equations},
  author={Tan, {Wei-Yan}},
  year={1992},
  address={Amsterdam, Netherlands},
  publisher={Elsevier}
}

@article {Tuleya:2007,
      author  = {Robert E Tuleya and Mark DeMaria and Robert J Kuligowski},
      title   = {Evaluation of {GFDL} and Simple Statistical Model Rainfall Forecasts for {U.S.} Landfalling Tropical Storms},
      journal = {Weather and Forecasting},
      year    = {2007},
      volume  = {22},
      number  = {1},
      pages   = {56--70},
      month   = {},
      note    = {},
      key     = {}}

@article{hope2013hindcast,
  title={Hindcast and validation of {Hurricane Ike (2008)} waves, forerunner, and storm surge},
  author={Hope, Mark E and Westerink, Joannes J and Kennedy, Andrew B and Kerr, PC and Dietrich, J Casey and Dawson, C and Bender, Christopher J and Smith, JM and Jensen, Robert E and Zijlema, Marcel and others},
  journal={Journal of Geophysical Research: Oceans},
  volume={118},
  number={9},
  pages={4424--4460},
  year={2013},
  publisher={Wiley Online Library}
}

@article{wichitrnithed2024discontinuous,
  title={A discontinuous {Galerkin} finite element model for compound flood simulations},
  author={Wichitrnithed, Chayanon and Valseth, Eirik and Kubatko, Ethan J and Kang, Younghun and Hudson, Mackenzie and Dawson, Clint},
  journal={Computer Methods in Applied Mechanics and Engineering},
  volume={420},
  pages={116707},
  year={2024},
  publisher={Elsevier}
}

@article{cangialosi2018tropical,
  title={Tropical cyclone report: {Hurricane Irma (AL112017)}},
  author={Cangialosi, John P and Latto, Andrew S and Berg, Robbie},
  journal={National Hurricane Center},
  volume={28},
  pages={2020},
  year={2018}
}

@article{callaghan2020extreme,
  title={Extreme rainfall and flooding from {Hurricane Florence}},
  author={Callaghan, Jeff},
  journal={Tropical Cyclone Research and Review},
  volume={9},
  number={3},
  pages={172--177},
  year={2020},
  publisher={Elsevier}
}

@article{pachev2023one,
  title={One-way coupling of {E3SM} with {ADCIRC} demonstrated on {Hurricane Harvey}},
  author={Pachev, Benjamin and Leung, L Ruby and Zhou, Tian and Dawson, Clint},
  journal={Natural Hazards},
  pages={1--25},
  year={2023},
  publisher={Springer}
}

@book{luettich2004formulation,
  title={Formulation and numerical implementation of the {{2D/3D ADCIRC}} finite element model version 44. XX},
  author={Luettich, Richard Albert and Westerink, Joannes J},
  volume={20},
  year={2004},
  publisher={R. Luettich},
  address={Chapel Hill, NC, USA}
}

@article{Manning1891,
  author  = {Manning, Robert},
  title   = {On the flow of water in open channels and pipes},
  journal = {Transactions of the Institution of Civil Engineers of Ireland},
  year    = {1891},
  volume  = {20},
  pages   = {161--207}
}

@article{Kennedy2011,
  title={Origin of the {Hurricane Ike} forerunner surge},
  author={Kennedy, Andrew B and Gravois, U and Zachry, BC and Westerink, JJ and Hope, ME and Dietrich, JC and Powell, MD and Atkinson, AT and Dean, RG and Bender, CJ},
  journal={Geophysical Research Letters},
  volume={38},
  number={8},
  year={2011},
  publisher={Wiley Online Library},
  doi={10.1029/2011GL047090}
}

@article{Rego2010,
  title={Nonlinear terms in storm surge predictions: Effect of tide and shelf geometry with case study from {Hurricane Rita}},
  author={Rego, Joao L and Li, Chunyan},
  journal={Journal of Geophysical Research: Oceans},
  volume={115},
  number={C6},
  year={2010},
  publisher={Wiley Online Library}
}

@article{sebastian2014,
  title={Characterizing hurricane storm surge behavior in {Galveston Bay using the SWAN+ADCIRC model}},
  author={Sebastian, Antonia and Proft, Jennifer and Dietrich, J Casey and Du, Wei and Bedient, Philip B and Dawson, Clint N},
  journal={Coastal Engineering},
  volume={88},
  pages={171--181},
  year={2014},
  publisher={Elsevier}
}

@article{ValleLevinson2020,
  title={Compound flooding in {Houston-Galveston Bay} during {Hurricane Harvey}},
  author={Valle-Levinson, Arnoldo and Olabarrieta, Maitane and Heilman, L},
  journal={Science of the Total Environment},
  volume={747},
  pages={141272},
  year={2020},
  publisher={Elsevier}
}

@article{Gori2020,
  title={The role of compound flood drivers in coastal inundation},
  author={Gori, Avantika and Lin, Ning and Smith, James},
  journal={Water Resources Research},
  volume={56},
  number={4},
  year={2020},
  publisher={Wiley Online Library}
}

@article{anderson2014,
  title={Variable response of coastal environments of the northwestern {Gulf of Mexico} to sea-level rise and climate change: Implications for future change},
  author={Anderson, John B and Wallace, Davin J and Simms, Alexander R and Rodriguez, Antonio B and Milliken, Kristy T},
  journal={Marine Geology},
  volume={352},
  pages={348--366},
  year={2014},
  publisher={Elsevier}
}

@techreport{Garrett2013,
  title={The deadliest, costliest, and most intense United States tropical cyclones from 1851 to 2006 (and other frequently requested hurricane facts)},
  author={Blake, Eric S and Rappaport, Edward N and Landsea, Christopher W and Miami, NHC},
  year={2007},
  institution={NOAA Technical Memorandum NWS TPC-5 NOAA/National Weather Service, National Centers for Environmental Prediction}
}

@book{Hayes1967,
  title={Hurricanes as Geological Agents: Case Studies of Hurricanes Carla, 1961, and Cindy, 1963},
  author={Hayes, Miles O.},
  year={1967},
  publisher={Bureau of Economic Geology, University of Texas at Austin},
  series={Report of Investigations No. 61},
  address={Austin, TX},
  note={The definitive study on storm surge sedimentation in Texas}
}

@article{monica20244500,
  title={4500-year paleohurricane record from the {Western Gulf of Mexico, Coastal Central TX, USA}},
  author={Monica, Sarah B and Wallace, Davin J and Wallace, Elizabeth J and Du, Xiaojing and Dee, Sylvia G and Anderson, John B},
  journal={Marine Geology},
  volume={473},
  pages={107303},
  year={2024},
  publisher={Elsevier}
}

@techreport{Latto2020,
  title={Tropical Cyclone Report: {Tropical Storm Imelda (AL112019)}},
  author={Latto, Andrew S and Berg, Robbie},
  institution={National Hurricane Center},
  year={2020},
  month={January},
  type={NOAA NWS Report},
  url={https://www.nhc.noaa.gov/data/tcr/AL112019_Imelda.pdf}
}

@techreport{Berg2021,
  title={Tropical Cyclone Report: {Tropical Storm Beta (AL222020)}},
  author={Berg, Robbie},
  institution={National Hurricane Center},
  year={2021},
  type={NOAA NWS Report},
  url={https://www.nhc.noaa.gov/data/tcr/AL222020_Beta.pdf}
}

@article{Moftakhari2017,
  title={Compounding effects of sea level rise and fluvial flooding},
  author={Moftakhari, Hamed R. and Salvadori, G. and AghaKouchak, A. and Sanders, B.F. and Matthew, R.A.},
  journal={Proceedings of the National Academy of Sciences (PNAS)},
  volume={114},
  number={37},
  pages={9785--9790},
  year={2017},
  publisher={National Academy of Sciences}
}

@article{Bakhtyar2020,
  title={A new 1D/2D coupled modeling approach for a riverine-estuarine system under storm events: Application to {Delaware River Basin}},
  author={Bakhtyar, R. and Maitaria, K. and Velissariou, P. and others},
  journal={Journal of Geophysical Research: Oceans},
  volume={125},
  number={8},
  pages={e2019JC015822},
  year={2020},
  publisher={American Geophysical Union},
  doi={10.1029/2019JC015822}
}

@article{Pena2022,
  title={Compound flood modeling framework for surface--subsurface water interactions},
  author={Peña, F. and Nardi, F. and Melesse, A. M.},
  journal={Natural Hazards and Earth System Sciences},
  volume={22},
  number={3},
  pages={775--796},
  year={2022},
  publisher={Copernicus Publications},
  doi={10.5194/nhess-22-775-2022}
}

@article{Ye2020,
  title={Simulating storm surge and compound flooding events with a creek-to-ocean model: Importance of baroclinic effects},
  author={Ye, F. and Zhang, Y. J. and Yu, H. and others},
  journal={Water Resources Research},
  volume={56},
  number={3},
  pages={e2019WR026063},
  year={2020},
  publisher={American Geophysical Union},
  doi={10.1029/2019WR026063}
}

@techreport{nhc2024beryl,
  author      = {John P. Cangialosi},
  title       = {Tropical Cyclone Report: Hurricane {Beryl} (AL022024)},
  institution = {National Hurricane Center},
  year        = {2025},
  month       = {January},
  url         = {https://www.nhc.noaa.gov/data/tcr/AL022024_Beryl.pdf}
}

@techreport{nws2024lch,
  author      = {{National Weather Service}},
  title       = {Hurricane {Beryl}: June 28-July 11, 2024},
  institution = {NOAA Lake Charles Weather Forecast Office},
  year        = {2024},
  url         = {https://www.weather.gov/lch/2024Beryl},
  note        = {Accessed: 2026-01-15}
}

@misc{nws2024hgx,
  author      = {{National Weather Service}},
  title       = {Hurricane {Beryl} - 2024: Post-Storm Summary},
  howpublished = {NOAA Houston/Galveston Weather Forecast Office},
  year        = {2024},
  url         = {https://www.weather.gov/hgx/beryl2024}
}

@techreport{hcfcd2024beryl,
  author      = {Jeff Lindner},
  title       = {Final Hurricane {Beryl} - July 8, 2024: Rainfall and Flood Information},
  institution = {Harris County Flood Control District},
  year        = {2024},
  month       = {August},
  url         = {https://reduceflooding.com/wp-content/uploads/2024/08/Hurricane-Beryl-Final-Report-Small.pdf}
}

@misc{noaa2024morgans,
  author      = {{NOAA Tides \& Currents}},
  title       = {Observed Water Levels: {Morgans Point, Barbours Cut, TX (Station 8770613)}},
  year        = {2024},
  month       = {July},
  url         = {https://tidesandcurrents.noaa.gov/stationhome.html?id=8770613},
  note        = {Verified peak water level of 5.45 ft MHHW on July 8, 2024}
}

@misc{usgs2024erosion,
  author      = {{U.S. Geological Survey}},
  title       = {Sixty-Eight Percent of {Texas} Coastline Likely to Experience Erosion Due to {Beryl}},
  year        = {2024},
  month       = {July},
  url         = {https://www.usgs.gov/news/state-news-release/sixty-eight-percent-texas-coastline-likely-experience-erosion-due-beryl}
}

@article{nabukulu2024,
  author    = {Nabukulu, Mary and Jetten, Victor G. and Ettema, Janneke},
  title     = {Implications of Tropical Cyclone Rainfall Spatial--Temporal Variability on Flood Hazard Assessments in the {Caribbean Lesser Antilles}},
  journal   = {GeoHazards},
  volume    = {5},
  number    = {4},
  pages     = {1275--1293},
  year      = {2024},
  month     = {November},
  publisher = {MDPI},
  doi       = {10.3390/geohazards5040060},
}

@techreport{ccrif2024,
  author      = {{CCRIF SPC}},
  title       = {Tropical Cyclone {Beryl} Event Briefing: {Windward Islands}},
  institution = {Caribbean Catastrophe Risk Insurance Facility},
  year        = {2024},
  month       = {July},
  url         = {https://www.ccrif.org/sites/default/files/publications/CCRIF_SPC_Event_Briefing_Tropical_Cyclone_Beryl_Windward_Islands_July_2024.pdf},
}

@article{matlock1982shoreline,
  title={Shoreline distances on the {Texas} coast},
  author={Matlock, Gary C and Ferguson, MFO and Areas, Shallow-water Surface},
  journal={Texas Parks and Wildlife Department, Coastal Fisheries Branch},
  year={1982}
}

@techreport{ipet,
  title  = {Performance Evaluation of the {New Orleans} and {Southeast Louisiana} Hurricane Protection System},
  volume = {Volume VIII – Engineering and Operational Risk
and Reliability Analysis},
  author = {{Interagency Performance Evaluation Task Force}},
  institution = {U.S. Army Corps of Engineers},
  year   = {2009}
}

@article{Li2020-llf,
  author = {Li, Jia and Zhang, Dazhi and Meng, Xiong and Wu, Boying and Zhang, Qiang},
  title = {Discontinuous {Galerkin} Methods for Nonlinear Scalar Conservation Laws: Generalized Local Lax--Friedrichs Numerical Fluxes},
  journal = {SIAM Journal on Numerical Analysis},
  volume = {58},
  number = {1},
  pages = {1-20},
  year = {2020},
  doi = {10.1137/19M1243798} 
}

@manual{floodeventviewer,
  author = {{U.S. Geological Survey}},
  institution = {U.S. Geological Survey database},
  title = {Short-Term Network Data Portal},
  year = {2026},
  url = {http://water.usgs.gov/floods/FEV},
  urldate = {2026-04-01}
}

@manual{usgstrespalacios,
  author = {{U.S. Geological Survey}},
  institution = {U.S. Geological Survey database},
  title = {USGS Current Conditions for the Nation},
  year = {2026},
  url = {https://nwis.waterdata.usgs.gov/nwis/uv?},
  urldate = {2026-08-24}
}

@manual{openstreetmap,
    author = {OpenStreetMap contributors},
    title = {OpenStreetMap [Data set]},
    institution = {OpenStreetMap Foundation},
    url = {https://openstreetmap.org},
    year = {2026}
}

\end{document}